\documentclass[a4paper,12pt]{article}
\usepackage{amsmath,amssymb,amsfonts,amsthm}
\newtheorem{theorem}{Theorem}[section]
\newtheorem{lemma}[theorem]{Lemma}
\newtheorem{definition}[theorem]{Definition}

\newtheorem{condition}[theorem]{Condition}
\usepackage{graphicx}
\usepackage{psfrag}

\newcommand{\dv}{\, d\mu}

\newcommand{\dvhc}{\, d\mu_{\tilde h}}

\newcommand{\ds}{\, ds}

\newcommand{\dshp}{\, ds_h}

\newcommand{\Rt}{\mathbb{R}^3}

\newcommand{\mf}{\mathcal{M}}

\newcommand{\vv}{\alpha}
\newcommand{\vY}{y}

\newcommand{\da}{\Gamma}
\newcommand{\dn}{A}

\newcommand{\cs}[1]{C^{\infty}_0(#1)}

\newcommand{\hx}{H^1_0(\Rt\setminus \{0\})}
\newcommand{\hy}{H^1_{0,X_0}(\Rt\setminus \da) }

\newcommand{\hxo}{H^1_0(\dn)}
\newcommand{\hyo}{H^1_{0,h}(\dn)}
 
\newcommand{\hxno}[1]{||#1||_{1;\dn}}
\newcommand{\hyno}[1]{||#1||_{1,h;\dn}}

\newcommand{\hxn}[1]{||#1||_{1}}
\newcommand{\hyn}[1]{||#1||_{1,X_0}}

\title{Proof of the  angular momentum-mass inequality for
  axisymmetric black holes} 
\author{Sergio Dain\\
 Facultad de Matem\'atica, Astronom\'{\i}a y F\'{\i}sica\\
Universidad Nacional de C\'ordoba \\
  Ciudad Universitaria (5000) C\'ordoba \\
  Argentina}

\begin{document}
\maketitle

\begin{abstract} 
  
  We prove that extreme Kerr initial data set is a unique absolute
  minimum of the total mass in a (physically relevant) class of
  vacuum, maximal, asymptotically flat, axisymmetric data for Einstein
  equations with fixed angular momentum. These data represent
  non-stationary, axially symmetric, black holes.
  
  As a consequence, we obtain that any data in this class satisfy the
  inequality $\sqrt{J} \leq m$, where $m$ and $J$ are the total mass
  and angular momentum of the spacetime.

\end{abstract}

\section{Introduction}
\label{sec:introduction}
 
An \emph{initial data set} for the Einstein vacuum equations is given by a
triple  $(S,   h_{ij},   K_{ij})$  where $S$
is a connected 3-dimensional manifold, $  h_{ij} $ a (positive
definite) Riemannian metric, and $  K_{ij}$ a symmetric tensor
field on 
$S$, such that the vacuum constraint
equations
\begin{align}
 \label{const1}
   D_j   K^{ij} -  D^i   K= 0,\\
 \label{const2}
   R -  K_{ij}   K^{ij}+  K^2=0,
\end{align}
are satisfied on $S$. Where $ {D}$ and $  R$ are the
Levi-Civita connection and the Ricci scalar associated with
$ {h}_{ij}$, and $  K =   K_{ij}   h^{ij}$. In
these equations the indices are moved with the metric $  h_{ij}$
and its inverse $  h^{ij}$.

The manifold $S$ is called \emph{Euclidean at infinity}, if there exists
a compact subset $\mathcal{K}$ of $S$ such that $S\setminus \mathcal{K}$ is the disjoint union of a
finite number of open sets $U_k$, and each $U_k$ is isometric to the
exterior of a ball in $\Rt$. Each open set $U_k$ is called an
\emph{end} of $S$. Consider one end $U$ and the canonical coordinates
$x^i$ in $\Rt$ which contains the exterior of the ball to which $U$ is
diffeomorphic. Set $r=\left( \sum (x^i)^2
\right)^{1/2}$. An initial  data set is called \emph{asymptotically
  flat} if $S$ is Euclidean at infinity, the metric   $ 
h_{ij}$  tends to the euclidean metric and $  K_{ij}$ tends to
zero as $r\to \infty$ in an appropriate way.  
These fall off
conditions (see \cite{Bartnik86} \cite{Chrusciel86} for the optimal
fall off rates) imply the existence of the total mass $m$ (or ADM mass
\cite{Arnowitt62}) defined at each end $U$ by
\begin{equation}
  \label{eq:30}
m=\frac{1}{16\pi}\lim_{r\to \infty} \oint_{\partial B_r} \left(\partial_j
    h_{ij}-\partial_i 
    h_{jj}\right ) n^i \ds,  
\end{equation}
where $\partial$ denotes partial derivatives with respect to $x^i$,
$B_r$ is the euclidean sphere $r=constant$ in $U$, $n^i$ is its
exterior unit normal and $\ds$ is the surface element with respect to the
euclidean metric. 

A central result concerning this physical quantity is the positive
mass theorem \cite{Schoen81} \cite{Witten81}:
\begin{equation}
  \label{eq:1}
m\geq 0, 
\end{equation}
for asymptotically flat, complete, vacuum, data; 
with equality only for flat data
(i.e; the data for Minkowski spacetime).

We will further assume that the data are \emph{axially symmetric}, which means
that there exists  a  Killing vector field  $\eta^i$, i.e;
\begin{equation}
  \label{eq:38}
 \pounds_\eta   h_{ij}=0,
\end{equation}
where $\pounds$ denotes the Lie derivative, 
which  has complete periodic orbits and such that
\begin{equation} 
  \label{eq:8b} 
 \pounds_\eta   K_{ij}=0.
 \end{equation} 

For axially symmetric data there exists another well defined physical
 quantity, namely the angular momentum $J$ associated with an
 arbitrary closed 2-surface $\Sigma$ in $S$ (the Komar integral of the
 Killing vector \cite{Komar59}, see also \cite{Wald84}).  
We define
 the angular momentum of $\Sigma$ by the following surface integral
\begin{equation} 
  \label{eq:69} 
  J(\Sigma)=\oint_\Sigma   \pi_{ij} \eta^i n^j \dshp, 
\end{equation} 
where $\pi_{ij}= K_{ij}- K h_{ij}$ and $n^i$,  $\dshp$ are,
respectively, the unit normal vector and the surface element with
respect to $h_{ij}$. As a consequence of equation \eqref{const1} and
the Killing equation \eqref{eq:38}, the vector $ \pi_{ij} \eta^j$ is
divergence free.  Then, by Gauss theorem, $J(\Sigma)=J(\Sigma')$ if
$\Sigma\cup \Sigma'$ is the boundary of a region contained in $S$
(i.e; $J$ depends only on the homology class of $S$). If $S=\Rt$ then
it follows that $J(\Sigma)=0$ for all $\Sigma$. In order to have non
zero $J$ the manifold $S$ must have a non trivial topology, for
example $S$ can have more than one end.

Let $\Sigma_\infty$ be any closed surface in a given end $U$ such that
it encloses the corresponding ball in $\Rt$. The total angular
momentum of the end $U$ is defined by $J\equiv J(\Sigma_\infty)$.

Physical arguments suggest the following inequality at any end
\begin{equation}
  \label{eq:70}
  m\geq \sqrt{|J|},
\end{equation} 
for any complete, asymptotically flat, axially symmetric and vacuum
initial data set (see \cite{Dain05e} and reference therein).
Moreover, the equality in \eqref{eq:70} should imply that the data set
is an slice of the extreme Kerr spacetime.

This inequality was proved for initial data set close to extreme Kerr
data set in \cite{Dain05d} \cite{Dain05e}.

The main result of this article is the following
\begin{theorem}
\label{t2}
Let $(  h_{ij},   K_{ij}, S)$ be a Brill data set (see
definition \ref{bd}) such that they satisfy condition \ref{cond}. 
 Then inequality \eqref{eq:70} holds. Moreover, the
equality in \eqref{eq:70} holds if and only if the data are a slice
of the extreme Kerr spacetime. 
\end{theorem}
Another way of stating this theorem is to say: \emph{extreme Kerr
  initial data is the unique absolute minimum among all Brill data set
  (which satisfies condition \ref{cond}) with fixed angular momentum.}

Let us discuss the hypotheses of this theorem. The first assumption is
that the data belong to the Brill class. This class of data is defined
in section \ref{sec:brill-data}, it involves certain technical
restrictions on both the topology of the manifold and the behavior of
the fields.  As it was mentioned above, theorem \ref{t2} is expected
to be true for general asymptotically flat, axisymmetric, vacuum,
complete data. Nevertheless, we emphasize that the Brill class
is physically relevant in the following sense: it contains the Kerr
black hole data, it also contains non stationary data (in particular
small deviations from Kerr) and gravitational radiation is not
constrained to be small in any sense.  In section \ref{sec:brill-data}
we review a well known procedure for constructing a rich class of
examples of this class of data set.

The second assumption, condition \ref{cond}, imply that the data have
non trivial angular momentum only at one end. The theorem is expected
to be valid without this restriction, however this generalization
appears to be quite difficult.

Theorem \ref{t2} generalizes the results presented in
\cite{Dain05d} \cite{Dain05e} in two ways. First, it does not involve
any smallness assumptions on the norm of the fields. In particular,
the data is not required to be close to extreme Kerr data. Second, the
Killing vector $\eta$ is not required to be hypersurface orthogonal.   

Theorem \ref{t2} will be a consequence of the following 
result in the calculus of variations. 
  
Let $\rho$ denote the cylindrical radius in $\Rt$ and $\da$ the axis
$\rho=0$.  Define 
\begin{equation}
  \label{eq:94}
h=2\log \rho.
\end{equation}
It is important to note that $h$ is
an harmonic function in $\Rt\setminus\da$.
Let $x,Y:\Rt\to \mathbb{R}$ be two arbitrary functions.  
 Consider the following functional 
\begin{equation}
  \label{eq:5c}
 \mf(x,Y)= \frac{1}{32\pi}\int_{\mathbb{R}^3}
  \left(|\partial  x |^2  + e^{-2x-2h} |\partial Y |^2  \right) \dv, 
\end{equation}
where $\dv$ is the volume element in $\Rt$ and the contractions are
with respect to the euclidean metric.
The relation between this functional and the mass of a Brill data set
is discussed in section \ref{sec:brill-data}, see also
\cite{Dain05c}.

The extreme Kerr initial data  $(x_0,Y_0)$ are  given by (see, for
example, \cite{Dain05d})
\begin{equation}
  \label{eq:79}
x_0=\log X_0-h, \quad Y_0 = \bar Y_0-
\frac{2J^2\cos\theta\sin^4\theta}{\Sigma},  
\end{equation}
where
\begin{equation}
  \label{eq:57b}
X_0  =\left(\tilde r^2+|J| +\frac{2|J|^{3/2} \tilde r \sin^2\theta
  }{\Sigma} \right)\sin^2\theta, \quad  \bar Y_0 =
  2J(\cos^3\theta-3\cos\theta),
\end{equation}
and
\begin{equation}
  \label{eq:58b}
\tilde r = r+\sqrt{|J|}, \quad 
\Sigma=\tilde r^2+|J| \cos^2 \theta.
\end{equation}
In these equations, $(r,\theta)$ are spherical coordinates in $\Rt$
(with $\rho=r\sin\theta$)
and $J$ is an arbitrary constant.

Let $\hx$ be the completion of $\cs{\Rt\setminus\{0\}}$ under the norm
\begin{equation}
  \label{eq:2}
\hxn{\vv}=\left( \int_{\Rt} |\partial \vv|^2 \dv\right)^{1/2},
\end{equation}
and $\hy$ the completion of $\cs{\Rt\setminus\da}$ under the norm
\begin{equation}
  \label{eq:2c}
\hyn{\vY}=\left( \int_{\Rt} X_0^{-2} |\partial \vY|^2 \dv\right)^{1/2}.
\end{equation}
We define the positive and negative part of a function $\vv$ by
$\vv^+=\max\{ \vv, 0\}$ and $\vv^-=\min\{ \vv, 0\}$.

\begin{theorem}
 \label{t1}
 Consider the functional $\mf$ defined by \eqref{eq:5c}.  Let $\vv\in
 \hx$, $\vY\in \hy$. Assume in addition that $\vv^-, yX_0^{-1} \in
 L^\infty(\Rt)$ and $ \vv , X_0^{-1}\vY\to 0$ as $r \to
 \infty$.  Then, the following inequality holds
\begin{equation}
  \label{eq:5}
\mf(x_0+\vv,Y_0+\vY)\geq \mf(x_0,Y_0),
\end{equation}
where $(x_0,Y_0)$ are the extreme Kerr data. 
Moreover, the equality in \eqref{eq:5} holds if and only if  $\vv=\vY=0$.
\end{theorem}
This theorem is a generalization of the results presented in
\cite{Dain05d} where a local version has been proved. 

Remarkably, $\vv$ and $\vY$ are not assumed to be axially symmetric in
this theorem (i.e, they can depend on the $\varphi$ coordinate).  
However, we emphasize that theorem \ref{t2} is only valid for axially
symmetric data (see the remark after theorem \ref{t:brill}).

It is important to note that for the extreme Kerr data the difference
$Y_0-\bar Y_0=y_0$ satisfies the hypothesis of theorem \ref{t1} (see
the appendix) Then inequality \eqref{eq:5} can be written in an
equivalent form
\begin{equation}
  \label{eq:81}
\mf(x_0+\vv,\bar Y_0+\vY)\geq \mf(x_0,Y_0).
\end{equation}
The function $\bar Y_0$ fixes the angular momentum of the data and it
also fixes the origin of coordinates. 

In theorem \ref{t1} we require the boundedness of the functions $\vv^-$
and $yX_0^{-1}$. It is possible to prove the same result without the
assumption on $\vv^-$ and with a stronger assumption on $y$, namely
$ye^{-h}\in L^\infty(\Rt)$ (see a previous version of this article in
\cite{Dain06c}).  The disadvantage of this choice is that the function
$y_0$ defined above does not satisfy this assumption: $y_0e^{-h}$ is
not bounded at the origin. And hence, important examples as
non-extreme Kerr and the
Bowen-York data (see section \ref{sec:brill-data}) are not included.
Also, without the assumption $\vv^-\in L^\infty(\Rt)$ the proofs are
more involved.  Nevertheless, I believe that for future generalization
of theorem \ref{t1} these arguments which do not make use of the
condition  $\vv^-\in L^\infty(\Rt)$ can be relevant. 

In section \ref{sec:global-minimum} we give an equivalent norm for the
Sobolev spaces $\hx$ and $\hy$. In particular, this shows the
equivalence between $\hx$ and the weighted Sobolev spaces studied in
\cite{Bartnik86}. 

\section{Brill data} 
\label{sec:brill-data}

The purpose of this section is to define a class of axially symmetric
initial data set, we will call it Brill class because it is inspired
in Brill's positive mass proof for axially symmetric data
\cite{Brill59}. The point in this definition is that in this class the
total mass satisfies the lower bound given by theorem \ref{t:brill}.

Axial symmetry implies certain local conditions on the fields $ 
h_{ij}$ and $  K_{ij}$. Let us consider first the metric $ 
h_{ij}$. For any axially symmetric metric there exists a coordinate
system $(\rho,z,\varphi)$ such the metric has locally the following
form
\begin{equation}
  \label{eq:44}
   h=  e^{(x-2q)}(d\rho^2+dz^2)+ \rho^2 e^x  (d\varphi + A_\rho
 d\rho+ A_z dz)^2,
\end{equation}
where the functions $x,q, A_\rho, A_z$ do not depend on $\varphi$.  In
these coordinates, the axial Killing vector is given by $\eta=
\partial/\partial \varphi$ and its norm is given by 
\begin{equation}
  \label{eq:83}
X=e^{x+h},
\end{equation}
where $h$ is given by \eqref{eq:94}. 

Let $  K_{ij}$ be a solution of equation
\eqref{const1} such that it satisfies \eqref{eq:8b}. 
Define the vector $  S^i$ by
\begin{equation}
  \label{eq:2t}
    S_i=  K_{ij}\eta^j-X^{-1}  \eta_i  
  K_{jk}\eta^j\eta^k, 
\end{equation} 
where $  \eta_i=  h_{ij}\eta^j$. Then, define
$K_i$ by
\begin{equation}
  \label{eq:73}
 K_i=   \epsilon_{ijk}  S^j \eta^k,
\end{equation}
where $  \epsilon_{ijk}$ is the volume element of $ 
h_{ij}$. Using equations \eqref{const1}, \eqref{eq:8b} and the Killing
equation \eqref{eq:38} we obtain
\begin{equation}
  \label{eq:73b}
  D_{[j}K_{i]}=0.
\end{equation}
Hence, there exists a scalar function $Y$ such that
\begin{equation}
  \label{eq:72}
K_i=\frac{1}{2}  D_i Y.
\end{equation}
Summarizing, axial symmetry implies that locally the metric has the
form \eqref{eq:44} and there exists a potential $Y$ for the second
fundamental form. 

\begin{definition}
\label{bd}  
We say that an initial data set $(  h_{ij},   K_{ij}, S)$
for the Einstein vacuum equations is a Brill data set if it satisfies
the following conditions.

  i) $S=\Rt\setminus \sum_{k=1}^N i_k $ where $i_k$ are points in
  $\Rt$ located at the axis $\rho=0$ of $\Rt$. 

  ii) The coordinates $(\rho,z,\varphi)$ form a global coordinate
  system on $S$ and the metric $ h_{ij}$ is given by \eqref{eq:44}.
  The functions $x,q, A_\rho, A_z$ are assumed to be smooth in $S$.
  The functions $x$ and $q$ satisfy
\begin{align}
  \label{eq:21}
x &=o(r^{-1/2}), \quad \partial x=o(r^{-3/2}), \\
q & =o(r^{-1}), \quad \partial q=o(r^{-2}),   \label{eq:21q}
\end{align}
as $r\to \infty$ and 
\begin{align}
  \label{eq:21b}
x & =o(r_{(k)}^{-1/2}), \quad\partial x=o(r_{(k)}^{-3/2}), \\   
q & =o(r_{(k)}^{-1}), \quad\partial q=o(r_{(k)}^{-2}), \label{eq:21bq}
\end{align}
as $r_{(k)}\to 0$. Where $r_{(k)}$ is the euclidean distance to the
end point $i_k$. 

Let $\da'$ be defined as $\da'=\da\setminus \sum_{k=1}^N i_k $. We assume
that
\begin{equation}
  \label{eq:9b}
  q|_{\da'}=0.
\end{equation}

iii) The second fundamental form satisfies
\begin{equation}
  \label{eq:78}
\pounds_\eta   K_{ij}=0, \quad   K=0.
\end{equation}
The corresponding potential $Y$  is a smooth function on
$S$ such that 
\begin{equation}
  \label{eq:77}
\int_{\Rt}| \partial Y|^2 e^{-2x-2h}\dv < \infty.
\end{equation} 
\end{definition}

Let us analyze the definition of Brill data.  Condition (i)
implies that $S$ is Euclidean at infinity with $N+1$ ends.  In effect,
for each $i_k$ take a small ball $B_k$ of radius $r_{(k)}$,  centered
at $i_k$, where $r_{(k)}$ is small enough such that $B_k$
 does not contain any other $i_{k'}$ with $k'\neq k$. Take $B_R$, with
large $R$, such that $B_R$ contains all points $i_k$. The compact set
$\mathcal{K}$ is given by $\mathcal{K}= B_R \setminus \sum_{k=1}^N B_k$ and the open sets
$U_k$ are given by $B_k\setminus i_k$, for $1 \leq k \leq N$, and
$U_0$ is given by $\Rt \setminus B_R$.  Our choice of coordinate makes
an artificial distinction between the end $U_0$ (which represent
$r\rightarrow \infty$) and the other ones.  This is convenient for our
purpose because we want to work always at one fixed end.  

The fall of conditions \eqref{eq:21}--~\eqref{eq:21q} imply that the
metric is asymptotically flat at the end $U_0$ (i.e, it satisfies the
conditions given in \cite{Bartnik86} \cite{Chrusciel86}). At the other
ends, the fall off conditions \eqref{eq:21b}--~\eqref{eq:21bq} are
more general, they include the standard asymptotically flat fall off
and they also include the fall off of the extreme Kerr initial data.
 
In a Brill data set there are  two geometrical scalar functions, the
norm of the Killing vector $X$ 
and the potential $Y$ which is related to the twist of the Killing vector
(also called the Ernst potential \cite{Ernst68}).
These scalars are well defined in the four dimensional spacetime which
results as the evolution of the data. In contrast, the function $x$
depends on a choice of coordinates on the data.

The total mass is essentially contained in the $1/r$ part of the
conformal factor $x$, due to our assumption on $q$.

The angular momentum is determined by the potential
$Y$ in the following way.  Define the intervals $I_k$, $0<k<N$, to be
the open sets in the axis between $i_k$ and $i_{k-1}$, we also define
$I_0$ and $I_N$ as $z< i_0$ and $z >i_N$ respectively. That is, $\da'=
\cup_{k=0}^N I_k$. 
 Since
$h$ is singular at the axis, the assumption \eqref{eq:77} implies that
the gradient $\partial Y$ must vanish at each $I_k$ and hence $Y$ is
constant at $I_k$. If $Y$ is a smooth function on $\Rt$ this of
course implies that $Y$ is constant at the whole axis. 
However, as we
will see, in order to have a non zero angular momentum $Y$ can not be
continuous at the end points $i_k$.

Let $\Sigma_k$ be a closed surface that encloses only the point
$i_k$. From equation \eqref{eq:69} we deduce
\begin{equation}
  \label{eq:45}
J_k \equiv J(\Sigma_k)=\frac{1}{8}\left (Y|_{I_k} -Y|_{I_{k-1}}\right ),
\end{equation}
where $J_k$ is the total angular momentum of the end $i_k$. The total angular
momentum of the end $r\to \infty$  is given by
\begin{equation}
  \label{eq:46}
J=\frac{1}{8}\left (Y|_{I_0} -Y|_{I_{N}}\right ),
\end{equation}
which is equivalent to 
\begin{equation}
  \label{eq:47}
  J=\sum_{k=1}^N J_k. 
\end{equation}

Finally, let us discuss the restrictions involved in the definition
\ref{bd} with respect to general asymptotically flat, axisymmetric,
complete and vacuum data. Locally, there is no restriction on the
metric and the only restriction on the second fundamental form is the
maximal condition $K=0$. Globally, we have assumed a particular
topology on the compact core $\mathcal{K}$ of the asymptotically flat
manifold $S$. Also, we have assumed that the metric has globally the
form \eqref{eq:44}.  The fall off conditions \eqref{eq:21} for $x$ is
a consequence of the standard definition of asymptotically flatness,
however the fall off conditions \eqref{eq:21q} for $q$ is an extra
assumption. Condition \eqref{eq:9b} for $q$ on the axis is a
consequence of the regularity of the metric at the axis, and hence it
is not a restriction.

The fundamental property of Brill data is the following.
\begin{theorem}
  \label{t:brill}
  The total mass  $m$  of a Brill data satisfies the following inequality
\begin{equation}
  \label{eq:66}
m\geq \mf(x,Y).
\end{equation} 
where $\mf(x,Y)$ is given by \eqref{eq:5c}.
\end{theorem}
This theorem extends Brill original proof \cite{Brill59} in two ways.
First, it allows for non zero $A$ in the metric \eqref{eq:44}, this
generalization was recently given in \cite{Gibbons06}, we use this
result in the following proof. The second extension is that the
topology of the data is non trivial, this was introduced in
\cite{Dain05c}. In particular this includes the topology of the Kerr
initial data.  It is important to recall that we are not introducing
any inner boundary, the mass is obtained as an integral over $S$, that
is, an integral over all the asymptotic regions (see the discussion in
\cite{Dain05c}).

\begin{proof}
Under our decay assumptions on $q$, we have that the total  mass of a
Brill data is given
by 
\begin{equation}
  \label{eq:84}
m=- \frac{1}{8\pi} \lim_{R\to \infty} \oint_{\partial B_R} \partial_r x \ds.
\end{equation}
The Ricci  scalar $  R$ of the metric $  h_{ij}$ is given by (see
\cite{Gibbons06})
\begin{equation}
  \label{eq:63}
-\frac{1}{8}  R e^{(x-2q)} = \frac{1}{4}\Delta x +\frac{1}{16}
|\partial x|^2 
- \frac{1}{4}\Delta_2 q   +\frac{1}{16}\rho^2 e^{2q} ( A_{\rho,z}-
A_{z,\rho})^2,  
\end{equation}
where $\Delta$ is the Laplacian in $\Rt$ and  $\Delta_2$ is the
2-dimensional Laplacian 
\begin{equation}
  \label{eq:85}
\Delta_2 q=q_{,\rho\rho}+q_{,zz}. 
\end{equation}
We want to integrate \eqref{eq:63} over $\Rt$. Let us analyze each
term individually. 
 Consider the first term in the
right hand side of \eqref{eq:63}. To perform the integral, we take
the compact domain $\mathcal{K}$ defined above, we have
\begin{equation}
  \label{eq:90}
  \int_{\mathcal{K}} \Delta x \dv= \int_{\partial\mathcal{K}} \frac{\partial x}{\partial n} \ds,
\end{equation} 
where $\partial / \partial n$ denotes normal derivative. The boundary $\partial \mathcal{K}$ is formed by the boundaries $\partial B_k$
and $\partial B_R$. Using the decay condition \eqref{eq:21b}, we get
that the contribution of $\partial B_k$ vanishes in the limit $r_{(k)}\to
0$. Using \eqref{eq:84}, we get that the contribution of  $\partial
B_R$ in the limit $R\to \infty$ is the mass.

Take the Ricci scalar in the left hand side of \eqref{eq:63}. We use the
hypothesis that the data have $  K=0$ and the constraint equation
\eqref{const2} to get
\begin{equation}
  \label{eq:91}
  R =   K_{ij}  K^{ij}.
\end{equation}
We will get a lower bound to the left hand side of \eqref{eq:91}. 
The metric \eqref{eq:44} can be written in the following form
\begin{equation}
  \label{eq:92}
  h_{ij}=q_{ij} + X^{-1} \eta_i \eta_j,
\end{equation}
where $q_{ij}$ is a positive definite metric in the orbit space. Using
this decomposition we get
\begin{equation}
  \label{eq:93}
  K_{ij}  K^{ij}=   K^{ij}  K^{kf}q_{ik}q_{jf}+
X^{-2} (  K^{ij}\eta_i \eta_j)^2+ 2X^{-1}  K^{ij} 
K^{kf} \eta_i \eta_k q_{jf}.
\end{equation}
The first two terms in the right hand side of this equation are
positive defined. Using the definitions \eqref{eq:2t} and
\eqref{eq:73}, the last term can be written as follows
\begin{align}
  \label{eq:95}
  K^{ij} 
K^{kf} \eta_i \eta_k
q_{jf} & =S^iS_i\\
& =\frac{1}{X}K^iK_i\\
&=\frac{1}{4X}  D^iY   D_iY\\
&=\frac{1}{4X}|\partial Y|^2 e^{-x+2q}.
\end{align} 
Then we get
\begin{equation}
  \label{eq:96}
  R e^{(x-2q)} \geq  \frac{1}{2X^2}|\partial Y|^2. 
\end{equation}

Take the term  $\Delta_2 q$ in \eqref{eq:63}.  Let $K_\delta$ be the
cylinder $\rho \leq \delta$ and consider the following domain
$A_\delta =\mathcal{K}\setminus K_\delta$.  We integrate over $A_\delta$ and
then take the limit $\delta \to 0$. The integral over  $A_\delta$ can
be written in the following form 
\begin{align}
  \label{eq:86}
\int_{A_\delta} \Delta_2 q \dv  &= 4\pi \int_{A_\delta} d \rho 
 \, d z \, (q_{,\rho \rho }+q_{,z z})  \rho, \\
  & = 4\pi \int_{A_\delta} d \rho \, d z \, \left(
    (\rho q_{,\rho} - q)_{,\rho }+ (\rho q_{,z})_{,z} \right).
\end{align}
We use the divergence theorem in two dimensions to transform this
volume integral in a boundary integral, that is
\begin{equation}
  \label{eq:87}
\int_{A_\delta} d \rho \, d z \, \left(
    (\rho q_{,\rho} - q)_{,\rho }+ (\rho q_{,z})_{,z} \right) =
  \oint_{\partial A_\delta} 
  \bar V\cdot \bar n \, d \bar s,  
\end{equation}
where $\bar n$ is the 2-dimensional unit normal, $ d \bar s$ the
line element of the 1-dimensional boundary and $\bar V$ is the
2-dimensional vector given in coordinates $(\rho, z)$ by
\begin{equation}
  \label{eq:88}
\bar V= ((\rho q_{,\rho} - q), (\rho q_{,z}) ).
\end{equation}
By \eqref{eq:9b} and the assumption that $q$ is smooth on $S$ (and hence
the derivatives $q_{,\rho}$ and $q_{,z}$ are bounded at $\da'$)  we have
that the vector $V$ vanishes at
$\da'$.  Then, using \eqref{eq:86} and \eqref{eq:87}  we get
\begin{equation}
  \label{eq:97}
\lim_{\delta \to 0} \int_{A_\delta} \Delta_2 q \dv  =\oint_{\partial
  \mathcal{K} } 
  \bar V\cdot \bar n \, d \bar s.
\end{equation} 
We take now the limit $R\to \infty$ and $r_{(k)} \to 0$.  We use the decay
conditions \eqref{eq:21q} and \eqref{eq:21bq} to obtain
\begin{equation}
  \label{eq:89}
\int_{\Rt} \Delta_2 q \dv =0.
\end{equation}

Since the last term in \eqref{eq:63} is positive, collecting these
calculations  we get \eqref{eq:66}.
\end{proof}

Since the data should satisfy the constraint equations
\eqref{const1}--\eqref{const2}, it is not obvious that we can
construct non trivial examples of Brill data.  One can easily check
that Schwarzschild data in isotropic coordinates is in the Brill class.
Another explicit examples are Brill-Lindquist data and the Kerr black
hole data (i.e., Kerr data with parameters such that inequality
\eqref{eq:70} is satisfied), see \cite{Dain05c}.

Let us discuss a general procedure to construct a rich family of Brill
data. For simplicity, we will assume that $A=0$ in equation
\eqref{eq:44}. Consider the metric
\begin{equation}
  \label{eq:105}
\tilde  h_{ij} = e^{-2q}(d\rho^2+dz^2)+ \rho^2   d\varphi^2.
\end{equation}  
This metric will be used as a conformal background for the physical
metric $ h_{ij}$, that is $  h_{ij}=e^x \tilde h_{ij}$.
We will take $q$ in \eqref{eq:105}  and the potential $Y$ as given functions. 

We first discuss how to construct solutions of the momentum constraint
\eqref{const1} from an arbitrary potential $Y$ and how to prescribe
the angular momentum of the solution.
Consider the
following tensor
\begin{equation}
  \label{eq:42}
  \tilde K^{ij}= \frac{2}{\rho^2} \tilde S^{(i} \eta^{j)}, 
\end{equation}
where 
\begin{equation}
  \label{eq:s}
 \tilde S^i=\frac{1}{2\rho^2}\tilde \epsilon^{ijk}\eta_j\tilde D_k Y,
\end{equation}
and $\tilde \epsilon_{ijk}$ denotes the volume element with respect to
$\tilde h_{ij}$ and   $\tilde D$ is the connexion with respect to
$\tilde h_{ij}$. The indices of the
tilde quantities are moved with $\tilde h_{ij}$ and its inverse
$\tilde h^{ij}$. The tensor $\tilde K^{ij}$ is symmetric, trace free, and it satisfies (see,
for example, the appendix in \cite{Dain99b})
\begin{equation}
  \label{eq:dk}
\tilde   D_i\tilde K^{ij}=0.
\end{equation}
Hence, for an
arbitrary function $Y$ we get a solution of equation \eqref{eq:dk}
given by \eqref{eq:42}. This, essentially, provides a solution of the
momentum constraint \eqref{const1}.

To control the angular momentum of the data we will prescribe the
behavior of $Y$ near the axis in the following way.  Take spherical
coordinates $(r_{(k)}, \theta_{(k)})$ centered at the end point $i_k$
and consider the following function
\begin{equation}
  \label{eq:65}
\bar Y_k = 2J_k (\cos^3\theta_{(k)}-3\cos\theta_{(k)}),
\end{equation}
where $J_k$ are arbitrary constants. The normalization factor is
chosen to be consistent with equation 
\eqref{eq:45}. Define
\begin{equation}
  \label{eq:68}
\bar Y = \sum_{k=0}^N \bar Y_k. 
\end{equation} 
Let $Y=\bar Y+y$, where $y$ vanishes at the axis. Then, the angular
momentum of $Y$ at the ends $i_k$ is given by the free constants $J_k$
in $\bar Y$. 

We discuss now the conditions on the function $q$.   
Define the Yamabe number of $\tilde h_{ij}$ to be 
\begin{equation}
  \label{eq:80}
\lambda = \inf_{0\neq \varphi \in \cs{S}} \frac{\int_{\Rt}
 \left( 8\tilde D^i \varphi \tilde D_i \varphi     +\tilde
   R\varphi^2\right) \dvhc }{\int_{\Rt} \varphi ^6\dvhc }.
\end{equation}
In order to construct a
Brill data, the metric $\tilde h_{ij}$  should satisfy the condition $\lambda >0$,
as we will see in the following theorem.

\begin{theorem}
\label{t:exisbd}
Let $q \in \cs{S}$ such that $\lambda >0$ and let $Y=\bar Y + y$, where $\bar Y$
is given by \eqref{eq:68} and $y \in \cs{\Rt\setminus \da}$. 
Then, there exist a function $x$ such that 
\begin{equation}
  \label{eq:74}
 h_{ij} = e^x\tilde h_{ij}, \quad K_{ij} = e^{-x/2} \tilde K_{ij},
\end{equation}
define a Brill data set, where $\tilde h_{ij}$ is given by
\eqref{eq:105}  and $\tilde K_{ij}$ is given by
\eqref{eq:42}.   
\end{theorem}
This theorem was proved in \cite{Cantor79} and \cite{Cantor81b} (see
also the correction in \cite{Maxwell05} of this article). There exists
more general version of the theorem \cite{Choquet99}
\cite{Maxwell04}. 
We have assumed that the functions  involved have compact support in order to
simplify the 
assumptions, but  decay conditions are also possible. 

\begin{proof}[Sketch]
What follows is the rewriting of our setting in terms of the one used
in these references. To simplify the discussion, let us follow the
existence theorem in section VIII of \cite{Choquet99}. 
    
Define the function $\psi_0$ by
\begin{equation}
  \label{eq:51}
\psi_0= \sum_{k=1}^N 1+ \frac{1}{r_{(k)}}. 
\end{equation}
Consider the  metric defined by the following conformal rescaling
\begin{equation}
  \label{eq:75}
\hat h_{ij}= \psi_0^4 \tilde h_{ij}.
\end{equation}
One can easily check that this metric is asymptotically flat with
$N+1$ ends. Moreover, the Yamabe number of the metric $\hat h_{ij}$
is the same as the the one for $h_{ij}$ because, by construction, it is a
conformally invariant quantity. Then, $\hat h_{ij}$ is in the positive
Yamabe class. Hence, we can apply the above mentioned theorem to
conclude that there exists a 
solution of the Lichnerowicz equation
\begin{equation}
  \label{eq:43}
\hat D^i \hat D_i \psi -\frac{\hat R}{8} = \hat K^{ij}\hat  K_{ij}\psi^{-7},
\end{equation}
such that $\psi \to 1 $ at the end point $i_k$.
Where $\hat K_{ij}$ is given by $\hat K_{ij}=\psi_0^{-2}\tilde K_{ij}$ with
$\tilde K_{ij}$ given by \eqref{eq:42},  hat quantities are defined with
respect to the metric $\hat h_{ij}$ and the indices are moved with
 this metric and its inverse. 

Define $x$ to be $e^x=(\psi \psi_0)^4$.
Then, it follows, by the
standard conformal transformation formulas, that  
\eqref{eq:74} define a solution of the constraint equations
\eqref{const1}--\eqref{const2}. 

The singular part of $x$ is
given by $\psi_0$, at the end point $i_k$ we have
\begin{equation}
  \label{eq:76}
x=O(-4\log r_{(k)}), \quad \partial x=O(r^{-1}_{(k)}),
\end{equation}
which is consistent with \eqref{eq:21}. 
\end{proof}

It remains to show how to achieve the condition $\lambda >0$.
This is given by theorem 4.2 in \cite{Cantor81b}. Applying this theorem
to the present case we get (see also \cite{Murchadha93}).

\begin{theorem}
  Let $q^0 \in \cs{S}$ and set $q=Cq^0$, where $C$ is a constant.
  Then, for $C$ small enough we have $\lambda >0$.
\end{theorem}

A simple but non trivial choice for $q$ which satisfies $\lambda >0$
is $q= 0$. This gives conformally flat solutions for the constraint
equations.  This kind of solutions are extensively used in numerical
simulations for black hole collisions (see the review article
\cite{Cook00}). Two examples are the Bowen-York spinning data
\cite{Bowen80} and the data discussed in \cite{Dain02c}.

The definition of Brill data is tailored to the hypothesis of theorem
\ref{t:brill}. However, in order to proof theorem \ref{t1} we need to
impose more conditions.  More precisely we assume the following.
Define $y=Y-\bar Y_0$ and $\vv=x-x_0$ where $\bar Y_0$ and $x_0$ are
given by \eqref{eq:79}.
\begin{condition}
\label{cond}
We assume
$y \in \hy $ and  $\vv^-, X_0^{-1} y \in L^\infty(\Rt)$ and $X_0^{-1} y
\to 0 $ as $r\to \infty$.
\end{condition}
The conditions on $y$ imply that $y$ vanishes at the axis $\da$ and
hence there exists only one end with non trivial angular momentum. The
location of this end is fixed by the function $\bar Y_0$.
However, let us emphasize that the data can have extra ends as
long as they have zero angular momentum. 

We have also assumed that
$\vv^-\in L^\infty(\Rt)$.  This implies an extra restriction on the
behavior of $x$ near the ends. In definition \ref{bd} we have assumed
the fall off behavior \eqref{eq:21b}  of $x$ near the ends, on the
other hand for extreme Kerr we have $x_0=-2\ln
r+O(1)$ near $r\to 0$. A  relevant class of fall
condition that satisfies both  \eqref{eq:21b} and $\vv^-\in
L^\infty(\Rt)$ is given $x=-\beta \ln r +O(1)$ near  $r\to 0$, 
for $\beta\geq 2$. In particular, this includes the asymptotically
flat ends $\beta=4$ described in theorem \ref{t:exisbd} (see equation
\eqref{eq:76}). 

Let us discuss important examples of Brill data that satisfies
condition \ref{cond}. First, extreme Kerr data. In this case we have
$\vv=0$ and $y=y_0=Y_0-\bar Y_0$. In the appendix we prove that the
function $y_0$ satisfies the assumptions in \ref{cond}. Second,
non-extreme Kerr black hole data (for the explicit form of the
functions $X$ and $Y$ for these data see the appendix of
\cite{Dain05c}). These data are asymptotically flat at the end $r\to
0$ and hence, by the discussion above, we have  $\vv^-\in
L^\infty(\Rt)$. Using a similar computation as the one presented for
extreme Kerr in the appendix we conclude that the function $y$ also
satisfies \ref{cond}. Finally, another two examples  of Brill data
that satisfy condition \ref{cond} are the Bowen-York data for only one
spinning black hole (i.e.  $Y=\bar Y_0$ and $q=0$) and the data
constructed in \cite{Dain02c} in which $Y=Y_0$ and $q=0$.

\section{Global Minimum}
\label{sec:global-minimum}

The crucial property of the mass functional defined in \eqref{eq:5c}
is its relation to the energy of harmonic maps from $\Rt$ to the
hyperbolic plane $\mathbb{H}^2$: they differ by a boundary term. Let
$h$ be an arbitrary harmonic function on a domain $\Omega$ in $\Rt$.
Define the mass functional over $\Omega$ as
\begin{equation}
 \label{eq:3}
 \mf_{\Omega}= \frac{1}{32\pi}\int_{\Omega}
  \left(|\partial  x |^2  + e^{-2x-2h}  |\partial Y |^2
  \right) \dv. 
\end{equation}
Then, using that $h$ is harmonic, we find the following
identity
\begin{equation}
  \label{eq:18}
 \mf_{\Omega}= \mf'_{\Omega}- \oint_{\partial \Omega}  \frac{\partial
   h}{\partial n} (h+2x)\ds,  
\end{equation}
where $\mf'_{\Omega}$ is given by
\begin{equation} 
  \label{eq:19}
\mf'_{\Omega}=\frac{1}{32\pi}\int_{\Omega}
  \left(\frac{ |\partial  X |^2  +   |\partial Y |^2}{X^2}
  \right) \dv, 
\end{equation}
and we have  defined the function $X$  by
\begin{equation}
  \label{eq:26}
X=e^{h+x}.
\end{equation}
The functional $\mf'_{\Omega}$ defines an energy for  maps
$(X,Y):\Rt \to \mathbb{H}^2$ where $\mathbb{H}^2$ denotes the
hyperbolic plane $\{(X,Y): X>0\}$, equipped with the negative constant
curvature metric
\begin{equation}
  \label{eq:53}
ds^2= \frac{dX^2+dY^2}{X^2}. 
\end{equation}
The Euler-Lagrange equations for the energy $\mf'_{\Omega}$ are given by
\begin{align}
  \label{eq:ha1}
  \Delta \log X &= -\frac{|\partial Y|^2}{X^2},\\
\label{eq:ha2}
\Delta Y   & =
2 \frac{\partial Y\partial X}{X}. 
\end{align}
The solutions of
\eqref{eq:ha1}--\eqref{eq:ha2}, i.e, the critical points of
$\mf'_{\Omega}$, are called harmonic maps from $\Rt \to \mathbb{H}^2$.
Since $\mf_{\Omega}$ and $\mf'_{\Omega}$ differ only by a boundary
term, they have the same Euler-Lagrange equations. 

Harmonic maps have been intensively studied, in particular the
Dirichlet problem for target manifolds with negative curvature has
been solved \cite{Hamilton75} \cite{Schoen83} \cite{Schoen82}.
However, these results do not directly apply in our case because the
equations are singular at the axis. In effect, the function $X$
represents the norm of the Killing vector (see equation \eqref{eq:83})
which vanishes at $\da'$ and this function appears in the denominator
of equations \eqref{eq:ha1}--\eqref{eq:ha2}.  This singular behavior
implies that the energy of the harmonic map is infinite as it can be
seen from equation \eqref{eq:19}.

Solutions of equations \eqref{eq:ha1}--\eqref{eq:ha2}, with this type
of singular behavior at the axis, represent vacuum, stationary,
axially symmetric solutions of Einstein equations. This equivalence
was discovered by Carter \cite{Carter73} based in the work of Ernst
\cite{Ernst68}. The relation between the stationary, axially symmetric
equations and   harmonic maps was discovered much
later by Bunting (the original work by Bunting is unpublished, see
\cite{Carter85}). In General Relativity, equations
\eqref{eq:ha1}--\eqref{eq:ha2} are important because they play a
central role in the black hole equilibrium problem (see
\cite{Carter85} and the review articles \cite{Chrusciel96}
\cite{Carter99}). Motivated by this problem, G. Weinstein in a series
of articles, \cite{Weinstein90} \cite{Weinstein92} \cite{Weinstein94}
\cite{Weinstein95} \cite{Weinstein96} \cite{Weinstein96b} (see also
\cite{Li92}), studied the  Dirichlet problem for harmonic maps with
prescribed singularities of this type.
Weinstein work will be particularly relevant here, let us briefly
describe it.

Weinstein constructs solutions of \eqref{eq:ha1}--\eqref{eq:ha2}
which represent stationary, axially symmetric, black holes with
disconnected horizons. To prove the existence of such solutions, he
defines the energy $\mf_{\Omega}$, with an appropriate harmonic function
$h$. This energy play a role of  an auxiliary functional in order to
``regularize'' the singular energy $\mf'_{\Omega}$ of the harmonic map. The
solution is a minimum of  $\mf_\Omega$ and the existence is proved with
a direct variational method. 

Our problem is related: we have a solution of
\eqref{eq:ha1}--\eqref{eq:ha2} (i.e, the extreme Kerr solution given
by \eqref{eq:79}--\eqref{eq:57b}) and we want to prove that it is a
unique minimum of $\mf$. There exists, however, two important
differences with Weinstein work. 

The first one, which is a simplification, is that we do not want to
prove existence of solution. We already have an explicit solution, we
just want to prove that it is a minimum.

The second difference, which introduces
a difficulty, is that we deal with the \emph{extreme} Kerr
solution. Extreme means that $m=\sqrt{|J|}$, where $m$ is the mass and
$J$ the angular momentum of the black hole, this definition can be
also extended for multiple black holes (see
\cite{Weinstein94}). This is a degenerate limit for black hole
solutions, it  is excluded in the
hypotheses of Weinstein existence theorems. Hence, these results do
not directly apply to our case. 

The extreme limit presents important peculiarities with respect to the
non extreme cases.  Remarkably enough, in this case (and only in this
case) the functional $\mf$ is the mass of the black hole (see
\cite{Dain05c}). In the non extreme cases, the functional defined by
Weinstein is not the same as our definition \eqref{eq:5c}, the choice
of the harmonic function $h$ is different.  In particular, if we take
the extreme limit of the Weinstein functional for one Kerr black hole
we get zero and not the total mass. Perhaps, Weinstein functional
describes the interaction energy of multiple black holes and this is
related to the non zero force between them. The existence of this
force in the general case is an open question. This question is
relevant for the black hole uniqueness problem with disconnected
horizons.

Another peculiarity of the extreme case is that the relevant manifold
is complete without boundary, in the non extreme case the manifold has
an inner boundary: the horizon of the black hole (there is no horizon
in the extreme Kerr black hole).

Let us give the main ideas of the proof of theorem \ref{t1}. Theorems
\ref{t:ex} and \ref{t:uniq} establish that extreme Kerr is the unique
minimum in an annulus centered at the origin, with appropriate
boundary conditions. The choice of the domain is important to avoid
the singularity of the extreme Kerr solution at the origin (this is
the main technical difference with the non extreme case).  These two
theorems are analogous to Proposition 1 and Proposition 3 of
\cite{Weinstein92} and use similar techniques. The main idea in the
proof of theorem \ref{t:ex} is the a priori bounds found by
Weinstein.  In theorem \ref{t:gu} we prove a uniqueness result for
extreme Kerr in the whole domain $\Rt$ under appropriate decay
conditions. This theorem is interesting by
itself. Finally, to prove theorem \ref{t1}, we
cover $\Rt$ with annulus and use a density argument together with the
previous theorems. This argument will work because we know a priori
the solution in $\Rt$. This is an important point, in this theorem we
are not proving the existence of the extreme Kerr solution. Note that
in \cite{Weinstein92}, theorem 1, it is proved the existence of
solution for the non extreme cases, this proof requires the a priori
bounds given by proposition 2 which are not valid in the extreme case.

Let $B_R$ be a ball of radius $R$ in $\Rt$ centered at the origin. 
We define the annulus  $\dn= B_R
\setminus B_\epsilon$, where  $R>\epsilon> 0$ are two arbitrary constants. 
Let $\hxo$ be the standard Sobolev space on  $\dn$, that is, the closure of
$\cs{\dn}$ under the norm
\begin{equation}
  \label{eq:hxo}
  \hxno{\vv}=\left( \int_{\dn} |\partial \vv|^2 \dv\right)^{1/2}.
\end{equation}  
And define the weighted Sobolev space $\hyo$ to be the closure of
$\cs{\dn\setminus\da}$ under the norm
\begin{equation}
  \label{eq:hyo}
\hyno{\vY}=\left( \int_{\dn}e^{-2h} |\partial \vY|^2 \dv\right)^{1/2}.
\end{equation}
Since the function $x_0$ is smooth on $\dn$ the norm \eqref{eq:hyo} is
equivalent to the norm \eqref{eq:2c} restricted to $\dn$. 

\begin{theorem}
\label{t:ex}
Consider the functional
defined by \eqref{eq:3} on the annulus $\dn$, with $h=2\log\rho$. Let $x_0$ and
$Y_0$ be the extreme Kerr solution given by \eqref{eq:79}.  
Then, there exist
\begin{equation}
  \label{eq:27}
\vv_0 \in \hxo , \quad \vY_0 \in \hyo,
\end{equation}
such that 
\begin{equation}
\label{eq:am}
\mf_{\dn}(x_0+\vv,Y_0+\vY) \geq \mf_{\dn}(x_0+\vv_0,Y_0+\vY_0),
\end{equation}
for all $\vv\in \hxo$ and $y\in \hyo$. Moreover, the minimum $(\vv_0, y_0)$
satisfies
\begin{equation}
  \label{eq:bounded}
 \vv_0 \in L^\infty(\dn), \quad e^{-h}\vY_0 \in L^\infty(\dn), 
\end{equation}
and the functions
\begin{equation}
  \label{eq:XY}
 X=e^{h+x_0+\vv_0}, \quad  Y=Y_0+y_0,   
\end{equation}
define a harmonic map from  $(X,Y):\dn \setminus \da
\to \mathbb{H}^2$; that is, they satisfy equations 
\eqref{eq:ha1}--\eqref{eq:ha2} on $\dn\setminus\da$. 
\end{theorem}

\emph{Remark:} the choice of the domain  is important because the function
$x_0$ is not bounded at the origin. The proof fails if
the domain includes the origin.

\begin{proof}

 Define  
\begin{equation}
  \label{eq:6}
m_0 = \inf_{\vv\in \hxo,\, \vY \in \hyo } \mf_{\dn}(\vv,\vY).
\end{equation}
Since $\mf$ is bounded below, $m_0$ is finite. 
 Note that the functional $\mf_\dn$ is not bounded for arbitrary
  functions in $\hxo\times\hyo$.  
 
Let $(\vv_n, \vY_n)$ be  a minimizing sequence, that is 
\begin{equation}
  \label{eq:4}
\mf_{\dn}(\vv_n,\vY_n) \rightarrow m_0 \text{ as } n \rightarrow \infty.
\end{equation}

To prove the existence of a minimum we will prove that there
exists some subsequence of $(\vv_n,\vY_n)$ which converges to an actual
minimizer $(\vv_0,\vY_0)$. To prove this, we will show that for every
minimizing sequence it is possible to construct another minimizing
sequence such that $\vv_n$ is uniformly bounded. Then, the existence
of a convergent subsequence follows from standard arguments (see
\cite{Weinstein90}).

We define $x_n,Y_n$ by
\begin{equation}
  \label{eq:16}
x_n=x_0+\vv_n, \quad Y_n=Y_0+\vY_n. 
\end{equation}
We first obtain a lower bound for $x_n$. 
Let
\begin{equation}
  \label{eq:7}
C_1 =\min_{\partial \dn}x_0,
\end{equation}
the constant $C_1$ depends on $R$ and $\epsilon$, in particular
$C_1\to \infty$ as $\epsilon \to 0$ because $x_0$ is singular at the
origin. This is the reason why the proof fails if the domain includes
the origin. Given $(x_n,\vY_n)$, define a
new sequence $(x'_n, \vY_n)$ as $x_n'=\max\{x_n,C_1\}$. Then one
can check that $\mf(\vv'_n,\vY_n) \leq \mf(\vv_n,\vY_n)$. Moreover,
$\vv'_n \in \hxo$.  This gives lower bounds for $\vv'_n$ on $\dn$
\begin{equation}
  \label{eq:29}
\vv'_n \geq C_1-x_0 \geq C_1-\max_{\dn}x_0=C'_1. 
\end{equation}
Using this lower bound, we want to prove that the minimizing sequence
can be chosen such that $\vv_n\in \cs{A}$ and $\vY_n\in
\cs{A\setminus\da}$. This is an important step in the proof, it
will be used in the following to calculate boundary integrals that
are not defined  for generic functions in $H^1$. Also, it plays an
essential role in the proof of theorem  \ref{t1}.

Define the set  $\mathcal{H}$ as the subset of $\hxo$ such that the
lower bound \eqref{eq:29} is satisfied. 
The functional $\mf_\dn$ is bounded for all
functions $y\in \hyo$ and $\vv\in\mathcal{H}$.  
By definition, for every $\vv\in \hxo$ and $ \vY \in \hyo $ there
exists a sequence $\vv_n\in \cs{A}$ and $\vY_n\in
\cs{A\setminus\da}$ such that $\vv_n\to \vv $ and $\vY_n\to \vY$ as
$n \to \infty$ in the norms \eqref{eq:hxo} and \eqref{eq:hyo}
respectively. If $\vv \in \mathcal{H}$, then by lemma \ref{lb}, we can
take $\vv_n$ such that $\vv_n \in \mathcal{H}$ for all $n$. 
For such sequence, we claim that
\begin{equation}
  \label{eq:117}
\lim_{n\to \infty}  \mf_{\dn}(\vv_n,\vY_n)= \mf_{\dn}(\vv,\vY). 
\end{equation}
To prove this we compute
\begin{equation}
  \label{eq:118}
 |\mf_{\dn}(\vv_n,\vY_n)- \mf_{\dn}(\vv,\vY)|\leq I_1+ I_2,
\end{equation}
where
\begin{align}
  \label{eq:119}
I_1 &= \frac{1}{32\pi}\int_{\dn}
  \left||\partial  x_n |^2 -|\partial  x |^2
  \right| \dv,\\
I_2 &= \frac{1}{32\pi} \int_{\dn}
  e^{-2h}\left|e^{-2x_n}  |\partial Y_n |^2-e^{-2x}  |\partial Y |^2
  \right| \dv.   \label{eq:119b}
\end{align}
For $I_1$ we have
\begin{align}
  \label{eq:120}
I_1 &=  \frac{1}{32\pi}\int_{\dn}
  \left|\partial  (x_n  +  x )\cdot \partial  (x_n  -  x )
  \right| \dv,\\
&\leq  \frac{1}{\sqrt{32\pi}}\left(\mf^{1/2}_{\dn}(\vv_n,\vY_n) +
  \mf^{1/2}_{\dn}(\vv,\vY) 
\right)\hxno{\vv-\vv_n}, \label{eq:120b}   
\end{align}
where in the last line we have used H\"older inequality.  The first
factor in the right hand side of \eqref{eq:120b} is bounded for all
$n$ and $\vv_n\to \vv$ in $\hxo$ then we obtain that $I_1\to 0$ as $n
\to \infty$.

A similar computation for $I_2$ leads to 
\begin{align}
  \label{eq:121}
I_2 & = \frac{1}{32\pi}\int_{\dn}
  e^{-2h}\left| \left(e^{-x_n} \partial Y_n +e^{-x} \partial Y 
  \right) \cdot \left(e^{-x_n} \partial Y_n -e^{-x} \partial Y 
  \right) \right| \dv \\
&\leq \frac{1}{\sqrt{32\pi}} \left(\mf^{1/2}_{\dn}(\vv_n,\vY_n) +
   \mf^{1/2}_{\dn}(\vv,\vY) 
\right) \left(I_{2,1}+I_{2,2}\right),  \label{eq:intu}
\end{align}
where
\begin{align}
  \label{eq:122}
I_{2,1}& =\left(\int_{\dn}
  e^{-2h-2x_0} |\partial Y|^2\left|e^{-\vv_n} -e^{-\vv}\right|^2
  \dv\right)^{1/2},  \\ 
I_{2,2}& = \left(\int_{\dn}
  e^{-2h-2x_0-2\vv_n} |\partial (y-y_n)|^2  \dv\right)^{1/2} 
\label{eq:122c}.
\end{align}
The function $x_0$ is positive on $\dn$, then it can be trivially
bounded by $e^{-2x_0}\leq 1$ and hence suppressed from the definitions
of $I_{2,1}$ and $I_{2,2}$.   However, for later use in the proof of
theorem \ref{t1},  we keep it in equations \eqref{eq:122}--\eqref{eq:122c}. 

We have $\vv_n\in \mathcal{H}$, then the integrand in $I_{2,1}$ is
bounded by a summable function for all $n$. Since $\vv_n\to \vv$ a.e.
we can apply the dominated convergence theorem to conclude that
$I_{2,1} \to 0$ as $n\to \infty$. For $I_{2,2}$ we use again that
$\vv_n\in \mathcal{H}$ to bound the exponential factor $e^{-\vv_n}$
for all $n$ and then the assumption $y_n\to y$ in $\hyo$ to conclude
that $I_{2,2} \to 0$ as $n\to \infty$. Hence, we have proved
\eqref{eq:117}.

Let $\vv_k\in \hxo,\, \vY_k \in
\hyo$ be a minimizing sequence.  Let $\vv_{k,n}\in \cs{A}$ and
$\vY_{k,n}\in\cs{A\setminus\da} $ such that   $\vv_{k,n}\to \vv_k$ and
 $\vY_{k,n}\to \vY_k$ as $n\to \infty$. Then we have
\begin{equation}
  \label{eq:123}
 |\mf_{\dn}(\vv_{k,n},\vY_{k,n})-
 m_0|\leq |\mf_{\dn}(\vv_{k,n},\vY_{k,n})-\mf_{\dn}(\vv_{k},\vY_{k})|+
| \mf_{\dn}(\vv_{k},\vY_{k})   - m_0| 
\end{equation}
For an arbitrary $\epsilon$, by \eqref{eq:6}, there exists $k$ such
that 
\begin{equation}
  \label{eq:124}
| \mf_{\dn}(\vv_{k},\vY_{k})   - m_0| \leq \epsilon/2.
\end{equation}
For this $k$, by \eqref{eq:117}, there exists $n$ such that
\begin{equation}
  |\mf_{\dn}(\vv_{k,n},\vY_{k,n})-\mf_{\dn}(\vv_{k},\vY_{k})| \leq \epsilon/2.
\end{equation}
Hence we conclude that
\begin{equation}
  \label{eq:6b}
m_0 = \inf_{k,n\in \mathbb{N} } \mf_{\dn}(\vv_{k,n},\vY_{k,n}).
\end{equation}

In order to obtain upper bounds, we exploit the symmetries of the
hyperbolic plane. Define the  following inversions
\begin{align}
  \label{eq:inv1}
\bar X &= \frac{X}{X^2+Y^2},\\
\bar Y &= \frac{Y}{X^2+Y^2}\label{eq:inv2}.
\end{align}
We have (see \cite{Weinstein90})
\begin{equation}
  \label{eq:9}
\frac{|\partial X|^2+|\partial Y|^2}{X^2}=\frac{|\partial \bar
  X|^2+|\partial\bar  Y|^2}{\bar X^2}.
\end{equation}
Let $\bar h$ be an arbitrary harmonic function, define $\bar x$ by 
\begin{equation}
  \label{eq:17}
\bar X = e^{\bar h + \bar x}.
\end{equation}
Using equations \eqref{eq:18} and \eqref{eq:9} we obtain the following
identity
\begin{equation}
  \label{eq:20}
\mf_{\dn}= \bar \mf_{\dn} + \oint_{\partial \dn}  
\left(\frac{\partial  \bar  h}{\partial n} (\bar h+2\bar x)-   \frac{\partial
  h}{\partial n} (h+2x)\right) \ds,  
\end{equation}
where $\bar \mf_{\dn}=\mf_{\dn}(\bar x, \bar Y)$. 

Take $h=\bar h$. Denote by $K_\delta$ the cylinder $\rho \leq \delta$. 
Since $h$ is singular on the axis, in order
to perform the integrals we will consider the domain $\dn_\delta=
\dn\setminus K_\delta$ for some small $\delta>0$ and then take the
limit $\delta \to 0$. The  boundary integral in
\eqref{eq:20} reduce to
\begin{equation}
  \label{eq:10}
C_\dn=\lim_{\delta \to 0} \oint_{\partial \dn_\delta} 2\frac{\partial
  h}{\partial n} (\bar x- x)\ds. 
\end{equation}
From \eqref{eq:inv1} and \eqref{eq:17} we deduce
\begin{equation}
  \label{eq:8}
\bar x- x = -\log (e^{2h+2x} + Y^2). 
\end{equation}
Then we have
\begin{equation}
  \label{eq:11}
\lim_{\rho\to 0} (\bar x- x) =-2 \log{|J|}, 
\end{equation}
where we have used that $\vY\in \cs{\dn\setminus\da}$ and 
$Y^2=Y^2_0=4J^2$ at $\da$.  We assume $J\neq 0$, the case $J=0$ is
trivial. Hence we obtain
\begin{equation}
  \label{eq:22}
\mf_{\dn}= \bar \mf_{\dn} +C_\dn,
\end{equation}
where
\begin{equation}
  \label{eq:23}
  C_\dn= -  16\pi (R-\epsilon) \log(4J^2) - \oint_{\partial \dn} 2\frac{\partial h}{\partial n} \log
  (e^{2h+2x_0} + Y_0^2) \ds.  
\end{equation}
The important point is that $C_\dn$  is finite. 

We can use the same argument as above to obtain lower bound for the
function $\bar x$ in $\dn$. Take
\begin{equation}
  \label{eq:12}
 C_2= \min_{\partial \dn}\bar x=  \min_{\partial \dn}\{ x_0- \log
(e^{2h+2x_0} + Y_0^2)\}. 
\end{equation}
As in the case of $C_1$, here we also have that $C_2 \to \infty$ as
$\epsilon \to 0$.  Note that  $C_2$ and $C_1$ are independently of
$\vv$ and $\vY$. 

As before, we can define a new function $\bar x' =
\max\{\bar x, C_2\}$, the energy of $\bar x'$ is less or equal the
energy of $\bar x$. Then $\bar x' \geq C_2$.  In the following 
we redefine $\bar x'$ by $\bar x$. From \eqref{eq:inv1} we have
\begin{equation}
  \label{eq:24}
\bar X\leq \frac{1}{X},  
\end{equation}
and then
\begin{equation}
  \label{eq:13}
e^{x}\leq e^{-2h-\bar x}\leq  e^{-2h-C_2},
\end{equation}
in $\dn$. Also, from \eqref{eq:inv1} we have
\begin{equation}
  \label{eq:25}
\bar X\leq \frac{X}{Y^2},  
\end{equation}
and then we deduce
\begin{equation}
  \label{eq:14}
Y^2 \leq  e^{-2h-2C_2}. 
\end{equation}

\begin{figure}
  \begin{center}
\psfrag{O}{$\dn$}
\psfrag{e}{$\epsilon$}
\psfrag{d}{$\delta$}
\psfrag{G}{$\Gamma$}
\psfrag{R}{$R$}
\psfrag{K+}{$K_+$}
\psfrag{K-}{$K_-$}
\psfrag{1}{$\partial^1$}
\psfrag{2}{$\partial^2$}

\resizebox{5cm}{!}{\includegraphics{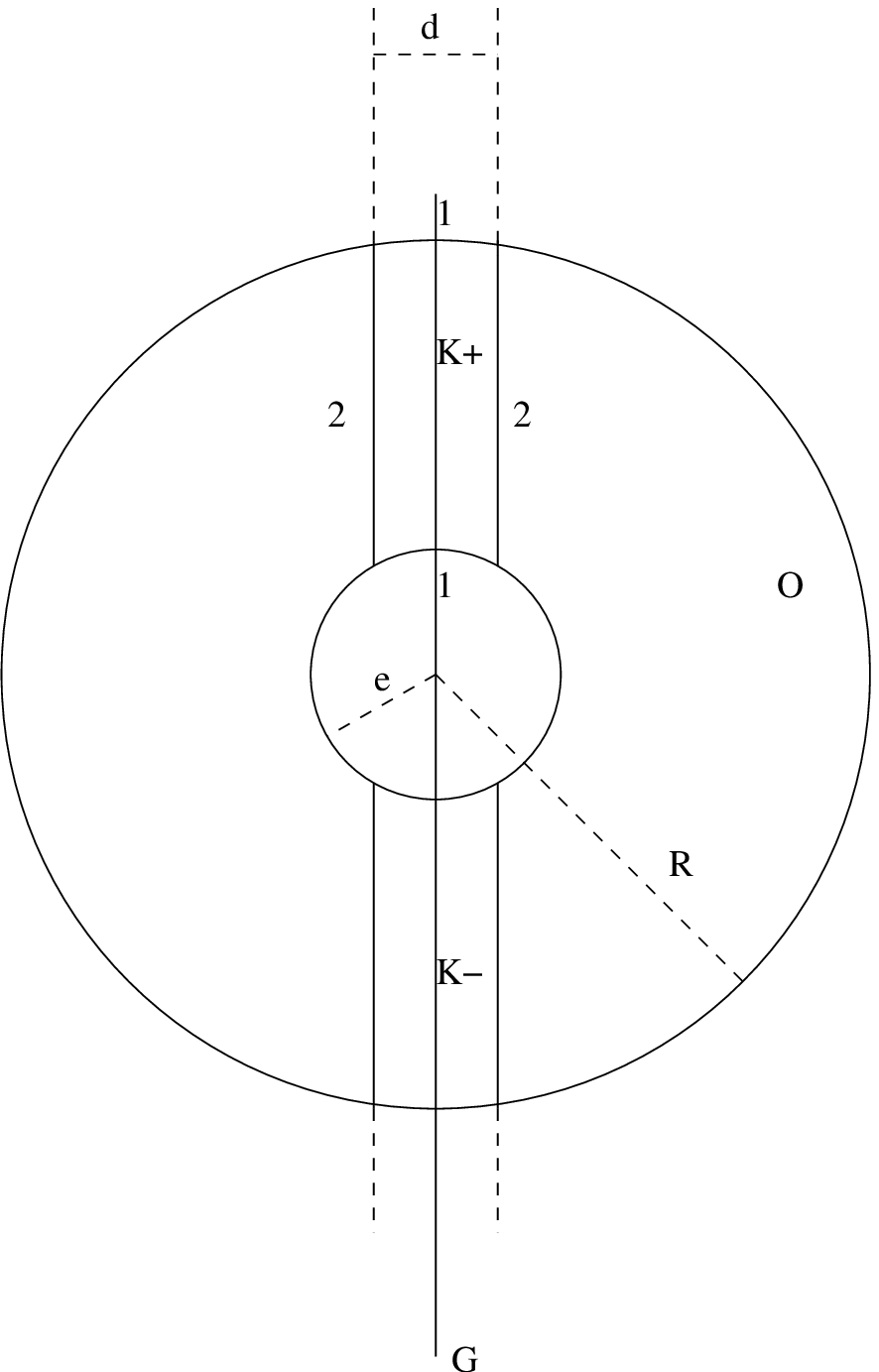}}
\caption{Domains}\label{fig:id}  
\end{center}
\end{figure}

We have obtained the bounds \eqref{eq:13} and \eqref{eq:14} which are
singular at the axis. To get bounds in a neighborhood of the axis we
will split this neighborhood in two disconnected domains: the upper
part and the lower one. More precisely, fix $\delta >0$ (we emphasize
that in this case we will not take the limit $\delta \to 0$ as
before), define $K_+= \dn \cap K_\delta \cap \{z\geq \epsilon \}$ and 
$K_-= \dn \cap K_\delta \cap \{z\leq \epsilon \}$, see figure \ref{fig:id}. 
 We will obtain estimates
for $K_+$ and $K_-$ independently.

On $K_+$ we define the following modified inversions
\begin{align}
  \label{eq:invJ1}
\bar X &= \frac{X}{X^2+(Y+2J)^2},\\
\bar Y &= \frac{Y}{X^2+(Y+2J)^2}\label{eq:invJ2}.
\end{align} 
Take $\bar h = -h$ and integrate
\eqref{eq:20} over $K_+$. The boundary term is given by 
\begin{equation}
  \label{eq:10c}
C_{K_+}=-2\oint_{\partial K_+} \frac{\partial h}{\partial n} (\bar x
+ x) \ds,
\end{equation}
where
\begin{equation}
  \label{eq:31}
\bar x = -\log \left(e^x+ e^{-2h-x}(Y +2J)^2\right). 
\end{equation}
We want to prove that $C_{K_+}$ is finite and the difficulty is of
course that $h$ is singular at $\da$.  We decompose the boundary
$\partial K_+$ into two pieces, the first one intersect the axis and
is given by $\partial^1 =\partial K_+ \cap \partial \dn$ and the
second does not intersect the axis and is given by
$\partial^2=\partial K_+ \cap \partial K_{\delta}$, see figure
\ref{fig:id}.  On $\partial^2$ the function $h$ is regular and hence
the integral is finite. On $\partial^1$ we have $\vY=\vv=0$. Using
that $y$ vanish near the axis and the following limit
\begin{equation}
  \label{eq:32}
\lim_{\rho \to 0} e^{-2h}(Y_0+2J)^2=0,
\end{equation}
we conclude that the integral is also finite in this piece of the
boundary. Equation \eqref{eq:32} is in fact the reason why in
equations \eqref{eq:invJ1}--\eqref{eq:invJ2} we have modified the
inversions \eqref{eq:inv1}--\eqref{eq:inv2} with the extra term $2J$.

 We can use now the same idea as before to obtain upper bounds. Set
\begin{equation}
  \label{eq:33}
C_3 =\min_{\partial K_+} \bar x =\min_{\partial K_+}
\{-\log(e^x+e^{-2h-x}(Y_0+2J)^2  \}.   
\end{equation}
By \eqref{eq:32} we have that this constant is finite. Then, we get
that $\bar x \geq C_3$ in $K_+$ and we can use the inversion to get
upper bounds for $x$ in $K_+$. However, here $C_3$ does depends on
$\vv$ and $\vY$ because these functions do not vanish on
$\partial^2$.  The key point is that nevertheless we can get lower
bounds to $C_3$ which does not depend on $\vv$ and $\vY$. In order
to do this we will use the previously defined constants
$C_2$ and $C_1$. The estimates are done in $\partial^1$ and $\partial^2$
independently. We decompose $C_3=C^1_3+C^2_3$ where 
\begin{align}
  \label{eq:34}
C^1_3 &=\min_{\partial^1} \{-\log(e^{x_0}+e^{-2h-x_0}(Y_0+2J)^2  \},\\
C^2_3 &=\min_{\partial^2} \{-\log(e^{x}+e^{-2h-x}(Y+2J)^2  \}.
\end{align} 
The constant $C^1_3$ does not depend on $\vv$
and $\vY$.  For $C^2_3$ we use the
previous estimate \eqref{eq:13} 
\begin{equation}
  \label{eq:35}
C^2_3  \geq \hat C^2_3,  
\end{equation}
where
\begin{equation}
  \label{eq:36}
\hat C^2_3  =- \log\left[\delta^{-4}\left( (e^{-C_2} +e^{-C_1}
    (2\delta^{-4}e^{-2C_2} +8J^2)
  \right) \right]. 
\end{equation}
does not depends on $\vv$ and $\vY$. Then, we conclude that $C_3\geq
C^1_3+\hat C^2_3$.
Hence, on $K_+$ we have
\begin{equation}
  \label{eq:37}
e^x\leq e^{-\bar x} \leq e^{-C_3} \leq e^{-(C^1_3+\hat C^2_3)},
\end{equation}  
and
\begin{equation}
  \label{eq:41}
(Y+2J)^2 \leq e^{-2C_3} e^{2h}\leq e^{-2(C^1_3+\hat C^2_3)  } e^{2h} .
\end{equation}
From \eqref{eq:41}, using $|a|-|b|\leq |a+b|$, we obtain that
$e^{-h}y$ is bounded.  Similar procedure can be used in $K_-$,
replacing $J$ by $-J$ in the inversions
\eqref{eq:invJ1}--\eqref{eq:invJ2}.
\end{proof}

We now turn to uniqueness. 
Let $(X_1,Y_1)$ and $(X_0,Y_0)$ be two points in $\mathbb{H}^2$. The
distance $d$ between these points in $\mathbb{H}^2$ is given by (see,
for example, \cite{Beardon83})
\begin{equation}
  \label{eq:49}
\cosh d = 1+\delta,
\end{equation}
where
\begin{equation}
  \label{eq:52}
\delta = \frac{1}{2} \frac{(X_1-X_0)^2  +(Y_1-Y_0)^2}{X_1X_0}.
\end{equation}
In our case,  $(X,Y)$ defines a map $(X,Y):\Rt\to\mathbb{H}^2 $, hence
 $d$ defines a function $d:\Rt\to\mathbb{R}$. Assume that $(X_1,Y_1)$
 and $(X_0,Y_0)$ are harmonic maps, we haven then the following
 two fundamental inequalities proved in \cite{Schoen79}
\begin{equation}
  \label{eq:61}
\Delta d^2 \geq 0, 
\end{equation}
and 
\begin{equation}
  \label{eq:54}
\Delta \sigma \geq 0, 
\end{equation}
where $\sigma= \sqrt{1+d^2}$.
These inequalities constitute  the basic ingredient in the uniqueness proof. 

Following \cite{Weinstein90}, we deduce from \eqref{eq:61}
\begin{equation}
  \label{eq:62}
\Delta \delta \geq 0,
\end{equation}
because $\delta$ is a convex function of $d^2$. 
Note that $\delta$ has a simpler expression in terms of $X,Y$ than
$d$.

Uniqueness proofs for the harmonic map equations
\eqref{eq:ha1}--\eqref{eq:ha2} constitute a fundamental step in the
black hole uniqueness theorems in General Relativity. The first result
in this subject was proved by Carter \cite{Carter71} at the linearized
level. Robinson \cite{Robinson75} obtain an identity for equations
\eqref{eq:ha1}--\eqref{eq:ha2} which lead to the first uniqueness
proof. The content of Robinson identity is essentially given by
\eqref{eq:62}. However, Robinson discover this identity independently
of \eqref{eq:61}. We emphasize that \eqref{eq:61} implies
\eqref{eq:62} but the converse is not true. 

In the context of black hole theory, \eqref{eq:61} is called the
Bunting identity  (see equation (6.48) in
\cite{Carter85}). This identity is not only more general than the
Robinson one but allow to extend the uniqueness proof to the charged
case.

The following uniqueness theorem in based on \eqref{eq:62}. 
\begin{theorem}
\label{t:uniq}
The solution founded in theorem \ref{t:ex} is unique and is given by $(0,0)$.
\end{theorem}

\begin{proof} 
  
  Let $(X_0,Y_0)$ be the extreme Kerr solution and let $(X_1,Y_1)$ be
  another solution of the harmonic map equations
  \eqref{eq:ha1}--\eqref{eq:ha2} on $\dn\setminus \da$, which
  satisfies \eqref{eq:XY}, \eqref{eq:27} and \eqref{eq:bounded}.
  
As usual, let  $x_0$ and $x_1$ be given by
\begin{equation}
\label{eq:39b}
X_0= e^{h+x_0} , \quad X_1=e^{h+x_1},
\end{equation} 
and define 
\begin{equation}
  \label{eq:39}
y=Y_1-Y_0, \quad   \vv= x_1-x_0.
\end{equation}
Let $\delta$ be given by \eqref{eq:52} and set 
\begin{equation}
  \label{eq:59}
\delta =\delta_x + \delta_y,
\end{equation}
where
\begin{equation}
  \label{eq:58}
\delta_x = \cosh\vv-1,
 \quad \delta_y= \frac{1}{2}  y^2e^{-2h-2x_0 -\vv},
\end{equation}
Note that by hypothesis $\delta=0$ on $\partial \dn$. 

Bellow, we will prove that $\delta\in H^1(A)$, let us assume that this
is true.  Since $\delta$ satisfies \eqref{eq:62} in $\dn\setminus \da$
we can apply lemma \ref{wr} to conclude that  \eqref{eq:62} is satisfied
 in $\dn$. Hence, we can use the weak maximum principle
for weak solutions (see \cite{Gilbarg}) in $\dn$. The function
$\delta$ is non negative in $\dn$ and it vanishes at the boundary,
then the weak maximum principle implies that $\delta=0$ in $\dn$ and
hence the conclusion follows.

It remains to prove that  $\delta\in H^1(\dn)$. In fact we will prove a
  stronger result: $\delta\in H^1(\dn)\cap L^\infty(\dn)$.  Recall that
  $x_0$ and $\vv$ are  bounded on $\dn$. Then, it follows that
  $\delta_x\in L^\infty(\dn)$. 
 From
  \eqref{eq:58} we get
\begin{equation}
  \label{eq:64}
\partial \delta_x = \sinh \alpha \partial \alpha,
\end{equation}
since $\vv \in H^1(\dn)$ it follows that $\delta_x\in H^1(\dn)$.

Consider $\delta_y$. Since $x_1$ and $e^{-h}y$ are bounded in $\dn$ we
conclude that $\delta_y\in L^\infty(\dn)$.  Its derivative is given by
\begin{equation}
\label{eq:ddelta}
\partial \delta_y= y \partial y e^{-2h-2x_0 -\vv}- y^2(\partial h + \partial x_0 +
\frac{1}{2}\partial \vv )e^{-2h-2x_0-\vv}. 
\end{equation}
Then, we have 
\begin{equation}
\label{eq:ddelta2}
|\partial \delta_y|^2\leq C \left( |\partial y|^2 e^{-2h} +
 (|\partial x_0|^2 +
\frac{1}{2}|\partial \vv |^2)- y^4e^{-4h}|\partial h|^2\right),
\end{equation}
where the constant $C$ depends only the $L^\infty$ norm of $\vv$,
$x_0$ and $ye^{-h}$.  When we perform the integral, the first three
terms are bounded since $y \in \hy$ and $\vv$, $x_0$ are in
$H^1(\dn)$. For the last term we use a Poincar\'e type inequality (see
lemma 1 of \cite{Weinstein92} and lemma 2.2  in \cite{Dain05d}).  We conclude
that $\delta_y$, and hence $\delta$, is in $H^1(\dn)\cap
L^\infty(\dn)$.
\end{proof}

\emph{Remark:} 
the proof of theorem \ref{t:uniq} fails if we extend to domain to
$\Rt$ because the function $\delta_x$ is not in $H^1(B_\epsilon)$.

In order to extend this theorem to $\Rt$ (or, in other words, in order
to generalize the uniqueness proofs to the extreme
cases) we will use inequality \eqref{eq:54} instead of \eqref{eq:61}
and \eqref{eq:62}. 

It is convenient to have an equivalent expression for $d$ in terms of
$\delta$. A straightforward computation gives 
\begin{equation}
  \label{eq:50}
d=2 \log (\sqrt{\delta}+ \sqrt{\delta+2} ) -\log 2,
\end{equation}
and hence the following expression for the derivative
\begin{equation}
  \label{eq:57}
\partial d = \frac{\partial \delta }{\sqrt{\delta(\delta+2)}}=
\frac{\partial \delta }{\sinh d}. 
\end{equation}
From \eqref{eq:50} we deduce the following important  inequalities
\begin{equation}
  \label{eq:60}
d\geq |\vv|,
\end{equation}
where $\vv$ is given by \eqref{eq:39} and
\begin{equation}
  \label{eq:28}
d \leq |\vv| + C,
\end{equation} 
where the constant $C$ depends only on the $L^\infty$ norm of
$\delta_y$ in $\Rt$.

Let us analyze the derivatives of $d^2$. Using \eqref{eq:60} and
\eqref{eq:57} we  obtain
\begin{equation}
  \label{eq:48}
|\partial d^2 |^2\leq 8d^2 |\partial \vv |^2+ 8d^2 |\partial \delta_y |^2.
\end{equation}
From this expression we get
\begin{equation}
  \label{eq:55}
|\partial \sigma|^2 \leq 2 \left(|\partial \vv|^2 + |\partial
  \delta_y|^2 \right). 
\end{equation}

Before proving theorem  \ref{t:gu}, we give an equivalent norm for the
relevant Sobolev spaces. 

Using a Poincar\'e type inequality (see Theorem 1.3 in
\cite{Bartnik86}), it follows that the norm \eqref{eq:2} on functions
in $\cs{\Rt\setminus\{0\}}$ is equivalent to the following weighted
norm
\begin{equation}
  \label{eq:2b}
\hxn{\vv}=\left( \int_{\Rt} |\partial \vv|^2 \dv\right)^{1/2}+\left(
\int_{\Rt}\frac{\vv^2}{r^2} \dv\right)^{1/2}.
\end{equation}
Then, the Sobolev space $\hx$ is equivalent to the weighted Sobolev
space $W'^{1,2}_{-1/2}$ studied in \cite{Bartnik86}.  In particular
from \eqref{eq:2b} we deduce that if $\vv\in \hx$ then $\vv\in
H^1_{loc}(\Rt)$. We also mention that the Sobolev inequality 
\begin{equation}
  \label{eq:112}
\left( \int_{\Rt} \vv^6 \dv \right)^{1/6}
\leq C \left(\int_{\Rt}|\partial  \vv|^2 \dv \right)^{1/2}, 
\end{equation}  
 is satisfied for all functions $\vv\in \hx$. 

Analogously, we can use
another type of  
Poincar\'e inequality 
 (see lemma  \ref{poincarex} ) to obtain an equivalent
norm to \eqref{eq:2c} for functions in $\cs{\Rt\setminus\da}$
 \begin{equation}
  \label{eq:2cequiv}
\hyn{\vY}=\left( \int_{\Rt}X_0^{-2} |\partial \vY|^2 \dv\right)^{1/2}+ \left(
\int_{\Rt}\frac{|\partial X_0|^2}{X_0^{4}} \vY^2 \dv\right)^{1/2}.
\end{equation}

\begin{theorem}[Uniqueness of extreme Kerr]
  \label{t:gu}
  Let $(X,Y)$ be a solution of the harmonic map equations
  \eqref{eq:ha1}--\eqref{eq:ha2} in $\Rt\setminus \da$. Define $(\vv,y)$
  by $X=e^{h+x}$, $Y=Y_0+y$, $x=x_0+\vv$. 
Assume that $\vv\in H^1_{loc}(\Rt)$, $y\in\hy$, 
  $yX_0^{-1},\vv^-\in L^\infty(\Rt)$ and that $\vv, yX_0^{-1}\to 0$ as
  $r\to \infty$. Then, $\vv=0$ and $y=0$.
\end{theorem}

\begin{proof}
 
  Let us analyze the function $\delta_y$ given by \eqref{eq:58}.  The
  computations are similar as in theorem \ref{t:uniq}, the difference is that here
  we have to take care of the singular behavior of the functions at
  the origin.
In
terms of $X_0$  the function $\delta_y$  is given by
\begin{equation}
  \label{eq:143}
  \delta_y=\frac{y^2e^{-\vv}}{2X_0^2}\leq\frac{y^2e^{-\vv^-}}{2X_0^2}. 
\end{equation}
Using the hypothesis  $yX_0^{-1},\vv^-\in L^\infty(\Rt)$ we obtain 
 $ \delta_y \in
  L^\infty(\Rt)$.

  Take a ball $B_R$ in $\Rt$ and consider the  the derivative of $\delta_y$
in $B_R$
\begin{equation}
  \label{eq:144}
\partial \delta_y= e^{-\vv}\left( \frac{y\partial y}{X_0^2}
  -\frac{y^2\partial \vv}{2X_0^2}-\frac{y^2\partial X_0}{X_0^3}\right).
\end{equation} 
Using our assumptions we conclude that the first to terms on the right
hand side of equation \eqref{eq:144} are in $L^2(B_R)$. For the third
term we use the assumption $yX_0^{-1}\in L^\infty(\Rt)$ and the Poincar\'e inequality given by lemma
\ref{poincarex}. Then, we conclude that $\delta_y$ is in $H^1(B_R)$.

Using inequality \eqref{eq:28} (which holds because we have proved
that $\delta_y$ is bounded) it follows that $\sigma\in L^2(B_R)$,
then using \eqref{eq:55} we obtain $\sigma\in H^1(B_R)$. Applying the
maximum principle to the inequality \eqref{eq:54}  we get
\begin{equation}
  \label{eq:56}
\sup_{\partial B_R } \sigma \geq \sup_{B_R } \sigma \geq 1.  
\end{equation}
Using the decay conditions we get that $\sup_{\partial B_R } \sigma
\to 1$ as $R\to \infty$. Then, it follows that $d=0$, and hence
$\vv=y=0$. 
\end{proof}

\begin{proof}[Proof of Theorem \ref{t1}] 

  We first prove the inequality \eqref{eq:5} using theorems \ref{t:ex}
  and \ref{t:uniq}.  The crucial step is to prove that the minimizing
  sequence can be chosen among functions with compact supports in
  annulus centered at the origin.

  Let $\vv\in \hx$ and $\vY\in \hy$. 
  By definition, there exist a sequence $y_n\in\cs{\Rt\setminus\da}$
  such that $y_n\to \vY$ in $\hy$ as $n\to \infty$. Let $R$ be the
  radius of a ball that contains the support of $y_n$. The radius
  $R$ depends on $n$ and we have that
  $R\to \infty$ as $n\to \infty$. For $\epsilon=1/R$, let
  $\chi_{\epsilon,R}$ be the cut off function defined in equation
  \eqref{eq:139} of the appendix.  Set $\vv_n=\vv
  \chi_{\epsilon,R}$, this function has compact support contained in
  the annulus $A_n=B_R\setminus B_\epsilon$ and $\vv_n\in
  H^1_0(A_n)$.  By lemma \ref{lbi} we have that $\vv_n\to \vv$ in
  $\hx$ as $n \to \infty$.  We claim that
\begin{equation}
  \label{eq:117b}
\lim_{n \to \infty}  \mf(\vv_n,\vY_n)=\mf(\vv,\vY).  
\end{equation}
This is similar to equation \eqref{eq:117} in the proof of theorem
\ref{t:ex}. Replacing the domain $\dn$ by $\Rt$, we define the same
integrals as in equations \eqref{eq:119}--\eqref{eq:119b}. Using
\eqref{eq:120}--\eqref{eq:120b} we conclude that $I_1\to 0$ as
$n \to \infty$. 

For the integrals $I_{2,1}$ and $I_{2,2}$ we use the hypothesis
$\vv^-\in L^\infty(\Rt)$ (which plays the same role as the lower bound
\eqref{eq:29} in the proof of theorem \ref{t:ex}) and 
\begin{equation}
  \label{eq:15}
  e^{-\vv_n}=e^{-\vv^+ \chi_{\epsilon,R}- \vv^- \chi_{\epsilon,R}}\leq
  e^{-\vv^- \chi_{\epsilon,R}}\leq e^{-\vv^-},
\end{equation}
to bound the terms with $e^{-\vv_n}$ by constants independent of $n$.
Using the assumption $y\in\hy$ we conclude that these two integrals
tend to zero as $n\to \infty$ and hence we have proved
\eqref{eq:117b}. 

Using a similar argument as in the proof of theorem \ref{t:ex}, from
equation \eqref{eq:117b} we conclude that the minimizing sequence
$(\vv_n, \vY_n)$  can
be taken among functions with compact support in annulus $A_n$.

 We  apply theorem \ref{t:ex} and theorem \ref{t:uniq} on $A_n$. We get
 \begin{equation}
   \label{eq:81b}
   \mf_{\dn_n}(x_0+\vv_n, Y_0+\vY_n) \geq  \mf_{\dn_n}(x_0, Y_0). 
 \end{equation}
Using this inequality we obtain 
 \begin{align}
   \label{eq:81c}
   \mf(x_0+\vv_n, Y_0+\vY_n) & = \mf_{\Rt\setminus \dn_n}(x_0, Y_0)+
   \mf_{\dn_n}(x_0+\vv_n, Y_0+\vY_n)\\
   & \geq  \mf_{\Rt\setminus \dn_n}(x_0, Y_0)+ \mf_{\dn_n}(x_0, Y_0)\\
   & = \mf(x_0, Y_0)\\
 & = \sqrt{|J|}.
 \end{align}
And then we get \eqref{eq:5}.

We prove now the rigidity part. Assume that there exist $\vv\in \hx$
and $\vY\in \hy$ such that
\begin{equation}
   \label{eq:82}
   \mf(x_0+\vv, Y_0+\vY) =  \mf(x_0, Y_0)=  \sqrt{|J|}.
 \end{equation}
 From inequality \eqref{eq:5} it follows that $(\vv,\vY)$ is a minimum
 of $\mf$, hence it satisfies the harmonic maps equations. 
 We use theorem \ref{t:gu} to conclude that $\vv=\vY=0$. 
\end{proof}

Finally, let us mention that theorem \ref{t2} follows directly from
theorem \ref{t1} and theorem \ref{t:brill}. Note that in the existence
proofs of section \ref{sec:brill-data} the free data are the functions
$q$ and $Y$, on the other hand in theorem \ref{t2} the free functions
are  $x$ and $Y$. Also, we emphasize that $x$ and $Y$ are not
necessarily axially symmetric in  \ref{t1}, however, the bound given by
theorem \ref{t:brill} require this condition. 

\section{Acknowledgments}

It is a pleasure to thank Piotr Chru\'sciel for illuminating
discussions and for a careful reading of an early version of this
article. These discussions began in the conference ICMP 2006, Rio de
Janeiro, Brazil. I would like to thank the organizers of this
conference for the invitation.

I would like also to thank the hospitality and support of the Isaac
Newton Institute of Mathematical Sciences in Cambridge, England. Most
of the discussions with P. Chru\'sciel and part of the writing of this
article took place during the program on ``Global Problems in
Mathematical Relativity'', October 2006.

The author is supported by CONICET (Argentina).
This work was supported in part by grant PIP 6354/05 of CONICET
(Argentina), grant 05/B270 of Secyt-UNC (Argentina) and the Partner
Group grant of the  Max Planck Institute for Gravitational Physics,
Albert-Einstein-Institute (Germany).   

\section{Appendix}

\begin{lemma}
\label{lb}
Let $\Omega$ be a bounded domain in $\mathbb{R}^n$ with $C^1$ boundary
$\partial \Omega$. Suppose that $u\in H^1_0(\Omega)$ and 
\begin{equation}
  \label{eq:67}
u\geq K,
\end{equation}
almost everywhere in $\Omega$, where $K\leq 0 $ is a  constant.  
Then, there exists a sequence
$u_n\in\cs{\Omega}$ such that
\begin{equation}
  \label{eq:98}
u_n\geq K,
\end{equation}
for all $n$ and $u_n \to u$ in the  $H^1_0(\Omega)$ norm.
\end{lemma}
\begin{proof}
The proof follows similar arguments as the proof of the trace zero
theorem for functions in $H^1_0(\Omega)$, see, for example, theorem 2
in chapter 5 of \cite{Evans98}. We will follow this
reference.  We will first prove the statement for functions in the half
plane which vanishes at the boundary and then we will extend this to
the domain $\Omega$. 

Let $(x',x_n)$ be coordinates in $\mathbb{R}^n$ and denote by
$\mathbb{R}^n_+$ the subset $x_n>0$. Let us assume that $u\in
H^1(\mathbb{R}^n)$, it has compact support in $\bar{\mathbb{R}}^n_+$
and vanishes on $\partial \mathbb{R}^n_+$.  Then, we can approximate
$u$ by smooth functions with compact support in $\bar{\mathbb{R}}^n_+$
which vanishes at the boundary $\partial \mathbb{R}^n_+$.
Integrating these functions and taking the limit to $u$ we obtain the
following estimate (see eq. (9), chapter 5, \cite{Evans98})
\begin{equation}
  \label{eq:133}
\int_{\mathbb{R}^{n-1}}|u(x',x_n)|^2\, dx'\leq C x_n \int_0^{x_n}
\int_{\mathbb{R}^{n-1}} |\partial u|^2\,dx'dt, 
\end{equation} 
for a. e. $x_n>0$.

Let $\chi:\mathbb{R}\to \mathbb{R}$ be a cut off function such that
$\chi \in C^\infty(\mathbb{R})$, $0\leq\chi\leq 1$, $\chi(t)=1$ for
$0\leq t \leq 1$, $\chi(t)=0$ for $2\leq t$ and $|d\chi/dt|\leq 1$ and
write $\chi_\epsilon(x)=\chi(x_n/\epsilon)$,
$u_\epsilon=(1-\chi_\epsilon) u$.  We want to prove that $u_\epsilon
\to u$ in $H^1(\Omega)$ as $\epsilon \to 0$. We have
\begin{equation}
  \label{eq:99}
||u_\epsilon-u ||^2_{L^2(\Omega)}=\int_{\Omega} u^2\chi^2_\epsilon \dv,
\end{equation}
since $u^2\chi^2_\epsilon\leq u^2$ (where, by hypothesis, $u^2$ is
measurable) and $u^2\chi^2_\epsilon \to 0$
a.e. as $\epsilon \to 0$  by the dominated convergence theorem we
conclude that the integral 
converges to zero as $\epsilon \to 0$. 
Consider the derivative
\begin{equation}
  \label{eq:100}
||\partial u_\epsilon-\partial u ||_{L^2(\Omega)}\leq  ||\chi_\epsilon  \partial u
||_{L^2(\Omega)}+ ||u \partial \chi_\epsilon
||_{L^2(\Omega)}.
\end{equation}
Using the same argument as above, we have that the first term in the
right hand side of this inequality goes to $0$ as
$\epsilon \to 0$. The delicate term is the second one. Note that the
derivative of $\chi_\epsilon$ has support in $\epsilon\leq x_n \leq
2\epsilon$ and that $|\partial \chi|\leq \epsilon^{-1}$, then we have
\begin{equation}
  \label{eq:134}
||u \partial \chi_\epsilon
||^2_{L^2(\Omega)}\leq
\epsilon^{-2}\int_\epsilon^{2\epsilon}\int_{\mathbb{R}^{n-1}} u^2 \, dx'dt.
\end{equation}
Using the estimate \eqref{eq:133} we obtain
\begin{align}
  \label{eq:135}
\epsilon^{-2}\int_\epsilon^{2\epsilon}\int_{\mathbb{R}^{n-1}} u^2 \,
dx'dt &\leq C \epsilon^{-2} \int_0^{2\epsilon} t\,
dt\int_\epsilon^{2\epsilon}\int_{\mathbb{R}^{n-1}}  |\partial u|^2 \,
dx'dx_n\\ 
& \leq C \int_\epsilon^{2\epsilon}\int_{\mathbb{R}^{n-1}} |\partial u|^2 \,
dx'dx_n,
\end{align}
and this integral tends to zero as $\epsilon \to 0$. Then we conclude 
\begin{equation}
  \label{eq:136}
u_\epsilon \to u \text{ in } H^1(\mathbb{R}^n_+).
\end{equation} 

Let $\eta_\delta$ be a mollifier. Since the functions $u_\epsilon$
have compact support in $\mathbb{R}^n_+$ we can mollify them to
construct smooth functions $u_{\epsilon,\delta}$ in
 $\mathbb{R}^n_+$.  Moreover, if $u$
satisfies the lower bound \eqref{eq:67} then $u_{\epsilon,\delta}$
satisfies it also. Indeed,
\begin{align}
  \label{eq:137}
u_{\epsilon,\delta}(x)=\int_{\mathbb{R}^n} \eta_\delta(x-y) u_{\epsilon}(y)\, dy &\geq K
\int_{\mathbb{R}^n} \eta_\delta(x-y)(1- \chi_{\epsilon})(y)\, dy \\
&\geq K,
\end{align}
where in the last line we have used that $K\leq 0$ and 
\begin{equation}
  \label{eq:138}
\int_{\mathbb{R}^n} \eta_\delta\, dx =1.
\end{equation}

To show that the functions $u_{\epsilon,\delta}$ converges to $u$ as
$\epsilon, \delta \to 0$ we write  
\begin{equation}
  \label{eq:114}
||u-u_{\epsilon,\delta} ||_{H^1}\leq ||u-u_{\epsilon} ||_{H^1}+
||u_{\epsilon} 
-u_{\epsilon,\delta} ||_{H^1},   
\end{equation}
and then  use that $u_{\epsilon,\delta}\to u_\epsilon$
as $\delta \to 0$ (this is the standard interior approximation in
$H^1$ by smooth functions, see for example, theorem 1, chapter 5, of
\cite{Evans98}) and that $u_{\epsilon}\to u$ as $\epsilon \to 0$.

We extend now this result to the domain $\Omega$ using a partition of
unity and flattering out the boundary.  Since $\partial \Omega$ is
compact, we can find finitely many points $x^0_i\in \partial \Omega$
and radii $r_i>0$, such that $\partial \Omega \subset \cup_{i=1}^N
B(x_i,r_i)$.  Define $V_i=\Omega\cap B(x_i,r_i)$ and let let
$V_0\subset \subset \Omega$, such that $\Omega \subset
\cup_{i=0}^NV_i$.

Let $\{\zeta \}^N_{i=0}$ be a smooth partition of unity of $\bar
\Omega$ subordinate to $V_i$. Define $u_i=u\zeta_i$, we have
\begin{equation}
  \label{eq:132}
u=\sum_{i=0}^Nu_i.
\end{equation} 
Consider $u_i$ for $i\geq 1$, since the boundary is $C^1$, it possible
to make a coordinate transformation such that it straightens out
$\partial \Omega$ near $x_i$. Then, we can assume that each $u_i$ has
compact support in $\bar{\mathbb{R}}^n_+$ and vanishes on $\partial
\mathbb{R}^n_+$. We use the result proved above to approximate each
$u_i$ by smooth functions with compact support which satisfy the lower
bound \eqref{eq:67}. Using \eqref{eq:132} we obtain the desired
conclusion. 
\end{proof}

The following function will be essential in the proofs of lemma
\ref{lbi} and \ref{wr}. It was taken from 
 \cite{Li92}, lemma 3.1.   Define
\begin{equation}
  \label{eq:106}
 t_\epsilon(\rho)=\frac{\log(-\log \rho)}{\log(-\log \epsilon)}
\end{equation}
and
\begin{equation}
  \label{eq:104}
\chi_\epsilon(\rho)=\chi(t_\epsilon(\rho)),
\end{equation}
where $\chi$ is the cut off function defined above.  The function
$t_\epsilon$ is defined for $0<\epsilon<1$ and $0<\rho<1$. We have that
$t_\epsilon\geq 2$ for $\rho \leq e^{(\log\epsilon)^2}$ and $0 \leq
t_\epsilon\leq 1$ for $\epsilon <\rho < e^{-1}$ (we assume $\epsilon$
small enough). It follows that the function $\chi_\epsilon$ defines a
smooth function in for $0\leq \rho  <\infty$ (we trivially extend the
function to be zero when $\rho \geq 1$). Moreover, $\chi_\epsilon(\rho)=0$
for $\rho\leq e^{-(\log\epsilon)^2}$ and $ \chi_\epsilon(\rho)=1$ for $r\geq
\epsilon$.

The derivative of $\chi_\epsilon$ has support in $
e^{-(\log\epsilon)^2}\leq\rho \leq \epsilon$ and is given by
\begin{equation}
  \label{eq:107}
\partial_\rho \chi_\epsilon= -\frac{d \chi_\epsilon
}{dt}\frac{1}{\log(-\log \epsilon)\rho\log \rho}.
\end{equation}
Assume $\epsilon \leq 1/2$, then we have
\begin{equation}
  \label{eq:108}
\int_0^\infty |\partial_\rho \chi_\epsilon|^2 \rho d\rho\leq
\frac{1}{(\log(-\log \epsilon))^2}\int_0^{1/2}  \frac{d\rho}{\rho(\log
  \rho)^2}. 
\end{equation}
The integral on the right hand side is bounded since
\begin{equation}
  \label{eq:111}
\int \frac{d\rho}{\rho(\log \rho)^2}=  -\frac{1}{\log \rho}.
\end{equation}
Then we obtain
\begin{equation}
  \label{eq:109}
\lim_{\epsilon\to 0}\int_0^\infty |\partial_\rho \chi_\epsilon|^2 \rho
d\rho=0.
\end{equation}
Take cylindrical coordinates $(\rho,z,\phi)$ in $\Rt$, the integral
\eqref{eq:109} is equivalent to 
\begin{equation}
  \label{eq:71}
\lim_{\epsilon\to 0}\int_0^\infty |\partial \chi_\epsilon|^2 \dv=0.
\end{equation}
This equation will be the crucial property of $\chi_\epsilon$ used in
the proof of lemma \ref{wr}. 

Consider now the spherical radius $r$, define $\chi_\epsilon(r)$ using the function $t_\epsilon(r)$ given
by \eqref{eq:106}. For $R>1$ we also define 
\begin{equation}
  \label{eq:106b}
 t_R(r)=\frac{\log(\log r)}{\log(\log R)},
\end{equation}
and
\begin{equation}
  \label{eq:104b}
\chi_R(r)=\chi(t_R(r)).
\end{equation}
Then, the following function has support in an annulus of radii
$e^{(\log R)^2}$ and $e^{-(\log \epsilon)^2}$
\begin{equation}
  \label{eq:139}
\chi_{\epsilon,R}(r)=\chi_R(r)+\chi_\epsilon(r)-1.
\end{equation}
A similar computation as above leads to 
\begin{equation}
  \label{eq:140}
\lim_{\substack{
\epsilon\to 0 \\
R\to \infty}} \int_{\Rt} |\partial
\chi_{\epsilon,R}|^3 \dv=0.
\end{equation}

\begin{lemma}
\label{lbi} 
Let $u\in \hx$. Then the functions $u_{\epsilon,R}=u\chi_{\epsilon,R}$
where $\chi_{\epsilon,R}$ is the cut off function defined in
\eqref{eq:139} converges to $u$ in the $\hx$ norm as $R\to \infty$,
$\epsilon \to 0$.  
\end{lemma}

\begin{proof}
We have
\begin{equation}
  \label{eq:100b}
||\partial u_{\epsilon,R}-\partial u ||_{L^2(\Rt)}\leq
||(1-\chi_{\epsilon,R})\partial u
||_{L^2(\Omega)}+ ||u \partial \chi_{\epsilon,R}||_{L^2(\Omega)}.
\end{equation}
The first term in the right hand side of this inequality goes to $0$
as $\epsilon \to 0$, $R\to \infty$.  For the second term we have 
\begin{align}
  \label{eq:101}
||u \partial \chi_{\epsilon,R} ||^2_{L^2(\Rt)} &\leq
||u^2||_{L^p(\Rt)}|||\partial\chi_{\epsilon,R}|^2 ||_{L^q(\Rt)}\\
&\leq ||\partial u||_{L^2(\Rt)} |||\partial\chi_{\epsilon,R}|^2
||_{L^{3/2}(\Rt)},
\end{align}
where in the first line we have used  
H\"older inequality with  $1/p+1/q=1$ and in the second line we chose 
$p=3$ and $q=3/2$ and use the Sobolev inequality \eqref{eq:112}. 
Then we use \eqref{eq:140} to obtain
the desired conclusion. 
\end{proof}

\begin{lemma}
\label{wr}  
Let $u\in H^1(\Omega)$ be a weak subsolution of the Laplace equation
  in $\Omega\setminus \da$. Then, $u$  is also a weak subsolution of
  the Laplace equation in $\Omega$. 
\end{lemma}
\begin{proof}
  By definition of weak subsolution in $\Omega\setminus \da$ we have
\begin{equation}
  \label{eq:103}
\int_{\Omega} \partial u \partial v  \dv\geq 0,
\end{equation}
for all $v \in \cs{\Omega \setminus \da}$. 
We want to prove that this inequality holds also for all 
$v \in \cs{\Omega}$.

Take cylindrical coordinates in $\Rt$ where $\rho$ is the distance to
the axis $\da$. Consider the cut off function $\chi_\epsilon(\rho)$
defined in \eqref{eq:104}.  
Let  $v \in
\cs{\Omega}$ and set $v=v(1-\chi_\epsilon) +  v\chi_\epsilon $.   Then
we have
\begin{equation}
  \label{eq:115}
\int_{\Omega} \partial u \partial v\dv=\int_{\Omega} \partial u \partial
(v(1-\chi_\epsilon))\dv+ \int_{\Omega} \partial u \partial
(v\chi_\epsilon))\dv \geq \int_{\Omega} \partial u \partial
(v(1-\chi_\epsilon)) \dv, 
\end{equation}
where we have used \eqref{eq:103} since $v\chi_\epsilon\in
\cs{\Omega \setminus \da}$.
We have
\begin{equation}
  \label{eq:141}
\int_{\Omega} \partial u \partial
(v(1-\chi_\epsilon))\dv\leq C ||u||_{H^1(\Omega)} ||\partial \chi_\epsilon
||_{L^2(\Rt)}. 
\end{equation}
We take the limit $\epsilon\to 0$ and use equation \eqref{eq:71} to
conclude that the integral goes to zero. Hence we conclude that
\begin{equation}
  \label{eq:110}
\int_{\Omega} \partial u \partial v\dv\geq 0,
\end{equation}
for all  $v \in
\cs{\Omega}$. 
  
\end{proof}

The following lemma gives a Poincar\'e type inequality for functions
in $\hy$.
\begin{lemma}
\label{poincarex}
Let $y\in \cs{\Rt\setminus\da}$ and $Y_0,X_0$ be given by \eqref{eq:57b}.
Then the following inequality holds
\begin{align}
  \label{eq:113}
\int_{\Rt}X_0^{-2} |\partial \vY|^2 \dv &\geq 
\int_{\Rt}\frac{(|\partial Y_0|^2+ |\partial X_0|^2)}{X_0^{4}} \vY^2
\dv\\ 
  &\geq 
\int_{\Rt}\frac{|\partial X_0|^2}{X_0^{4}} \vY^2 \dv.
\end{align}
\end{lemma}
\begin{proof}
  We use the following general identity proved in Proposition C.2 of
  \cite{Chrusciel03}
\begin{equation}
  \label{eq:142}
\int_{\Rt} e^{2v} |\partial y|^2 \dv\geq \int_{\Rt} e^{2v}(\Delta v +
|\partial v|^2)|y|^2\dv,
\end{equation}  
for $v=x_0+h$. Using equation \eqref{eq:ha1} the conclusion follows. 
\end{proof}
Finally, let us prove that the function
\begin{equation}
  \label{eq:40}
 y_0=Y_0-\bar Y_0 =-\frac{2J^2\cos\theta\sin^4\theta}{\Sigma},
\end{equation}
 defined in
 the introduction satisfies the hypothesis of theorem
\ref{t1}. Note that $y_0\in C^\infty(\Rt\setminus \{0\})$. Using
equation   \eqref{eq:57b} we obtain the lower bound
\begin{equation}
  \label{eq:102}
  X_0\geq |J| \sin^2\theta.
\end{equation}
Then we get
\begin{equation}
  \label{eq:116}
  \frac{|y_0|}{X_0}\leq \frac{2|J|}{(r+\sqrt{|J|})^2},
\end{equation}
which implies $|y_0|/X_0^{-1}\leq 2$ and  $|y_0|/X_0^{-1}\to 0$ as
$r\to \infty$. This bound also implies that
$y_0/X_0^{-1}\in L^p(\Rt)$ for $3/2<p$.  

Remains to show that $y_0\in \hy$. From \eqref{eq:40} we can
explicitly compute the norm \eqref{eq:2c} to prove that it is finite.
Take the sequence $y_{\epsilon,R}=y_0\chi_\epsilon(\rho)\chi_R(r)$
where $\chi_\epsilon(\rho)$ and $\chi_R(r)$ are given by
\eqref{eq:104} and \eqref{eq:104b}. We have that $y_{\epsilon,R}\in
\cs{\Rt\setminus \da}$. To prove that $y_{\epsilon,R}\to y$ in $\hy$
as $R\to \infty$, $\epsilon \to 0$, we use the same argument as in the
proof of lemma \ref{lbi} and the fact that $y_0/X_0^{-1}\in L^6(\Rt)$.


\end{document}